\documentclass[sigconf]{acmart}
\usepackage{multirow} % For merging rows
\usepackage{subcaption}
\usepackage{enumitem}
\usepackage{booktabs, tabularx}
\usepackage{xcolor}
\usepackage{makecell} % For multi-line and rotated headers

\AtBeginDocument{%
  }

\acmISBN{978-1-4503-XXXX-X/18/06}

\begin{document}

%\title{What You See is Not What You Searched for: Pixel Poisoning Attack Vulnerabilities of Vision-Language Model Retrievers}
%\title{Pixel Poisoning Attack Vulnerabilities of \\Vision-Language Model Retrievers}

\title{Document Screenshot Retrievers are Vulnerable to Pixel Poisoning Attacks}


\author{Shengyao Zhuang}
\authornote{Equal Contribution}
\affiliation{%
  \institution{CSIRO}
  \city{Brisbane}
  \state{QLD}
  \country{Australia}}

\author{Ekaterina Khramtsova}
\authornotemark[1]
\affiliation{%
	\institution{The University of Queensland}
	\city{Brisbane}
	\state{QLD}
	\country{Australia}}

\author{Xueguang Ma}
\affiliation{%
	\institution{University of Waterloo}
	\city{Waterloo}
	\country{Canada}}

\author{Bevan Koopman}
\affiliation{%
	\institution{CSIRO, The University of Queensland}
	\city{Brisbane}
	\state{QLD}
	\country{Australia}}

\author{Jimmy Lin}
\affiliation{%
	\institution{University of Waterloo}
	\city{Waterloo}
	\country{Canada}}

\author{Guido Zuccon}
\affiliation{%
	\institution{The University of Queensland}
	\city{Brisbane}
	\state{QLD}
	\country{Australia}}
%\email{jpkumquat@consortium.net}


\begin{abstract}
Recent advancements in dense retrieval have introduced vision-language model (VLM)-based retrievers, such as DSE and ColPali, which leverage document screenshots embedded as vectors to enable effective search and offer a simplified pipeline over traditional text-only methods.
In this study, we propose three pixel poisoning attack methods designed to compromise VLM-based retrievers and evaluate their effectiveness under various attack settings and parameter configurations. Our empirical results demonstrate that injecting even a single adversarial screenshot into the retrieval corpus can significantly disrupt search results, poisoning the top-10 retrieved documents for 41.9\% of queries in the case of DSE and 26.4\% for ColPali. These vulnerability rates notably exceed those observed with equivalent attacks on text-only retrievers. Moreover, when targeting a small set of known queries, the attack success rate raises, achieving complete success in certain cases.
By exposing the vulnerabilities inherent in vision-language models, this work highlights the potential risks associated with their deployment.

\end{abstract}





\maketitle

\section{Introduction}
%Large language model-based dense retrievers have demonstrated impressive retrieval effectiveness~\cite{}. However, these systems are vulnerable to a specific type of attack known as corpus poisoning~\cite{}, where malicious documents are injected into the index, compromising retrieval quality. On the other hand, 
Emerging vision-language models (VLMs) have brought new opportunities in document retrieval~\cite{faysse2024colpali,ma2024unifying}. VLM-based dense retrievers (e.g., DSE~\cite{ma2024unifying} and ColPali~\cite{faysse2024colpali}) embed document screenshots into vector representations and have proven effectiveness in retrieval tasks; they have the added advantage of simplifying the indexing pipeline by directly utilising visual features, removing the need for extra components like optical character recognition or table parsing methods.

%Despite recent advancements, 
In this paper we show that document screenshot retrievers powered by vision-language models (VLMs) are particularly susceptible to corpus poisoning and search engine optimisation (SEO) attacks. This is because, unlike text-based dense retrieval, where methods like Hotflip~\cite{zhong-etal-2023-poisoning,li2025reproducing} are typically used to generate adversarial documents for corpus poisoning, the image-based nature of VLMs introduces a novel attack vector. Specifically, the pixel values of document screenshots can be directly manipulated using gradients to deceive the model --- an attack strategy that has been extensively studied in computer vision but has not been considered in the context of document screenshot retrieval. %In this paper, we demonstrate that attack methods from the computer vision domain can be effectively adapted to target corpus poisoning and SEO in VLM-based document retrievers.

We develop three pixel-based attack methods for attacking document screenshot retrievers: 1) \textit{\textbf{Direct Optimisation}}, 2) \textit{\textbf{Noise Optimisation}}; and 3) \textit{\textbf{Mask Direct Optimisation}}. All methods begin with a seed document screenshot image, where the malicious attacker aims to optimise the ranking of this image for a target retriever and a group of queries. For the \textit{Direct Optimisation}, we calculate the gradient on image pixels to maximise the embedding similarity between the seed image and the target queries. The gradients are then directly used to update the image pixels. For the \textit{Noise Optimisation} method, we initialise a noise pixel matrix and add the noise to the image's original pixels. The gradient on the noise matrix is then calculated to optimise the similarity. For \textit{Mask Direct Optimisation}, we add a mask margin around the seed image and only update the pixels within the margin using the gradients. All three methods have parameters that control how many pixels will be updated, allowing for a trade-off between the fidelity of the adversarial image and the attack success rate.

We conducted experiments to evaluate the effectiveness of the attacks across varying levels of task difficulty: from targeting a small set of known (seen) queries, to targeting unseen queries from the same distribution as the training data, and finally, to targeting unseen queries from out-of-domain distributions. 
%on the corpus poisoning task in both in-domain and zero-shot settings, as well as the SEO task. 
Our results reveal that VLM-based dense retrievers are particularly vulnerable to our proposed pixel-based attack. For example, our results demonstrate that for in-domain attacks, even injecting a single adversarial screenshot document, generated with our attack methods, into the retrieval corpus can successfully poison the top-10 retrieved documents for 41.9\% of queries in the case of DSE and 26.4\% for ColPali. The success rate dramatically increases when attacks target a small set of known queries, where complete success can be reached even when performing small manipulations to the adversarial document. These findings have practical implications for the deployment of VLM-based dense retrievers, as such attacks can be exploited for corpus poisoning and SEO manipulation.

\section{Related Work}

\subsection{Document screenshot retrievers}
Dense retrievers, which encode both documents and queries into vector embeddings and estimate relevance based on the similarities between these embeddings, have demonstrated strong semantic matching abilities~\cite{zhao2024dense,luan2021sparse}. Most prior work focuses on pure text-based query and document representations, where the base encoder models are typically text-based transformers. 

Recently, a novel vision-based dense retrieval paradigm has em\-erged. Methods here represent documents using screenshot images and employs vision-language models to encode text queries and screenshot-based documents into embeddings. Compared to pure text-based dense retrieval pipelines, these methods eliminate the need for complex document preprocessing steps, such as table parsing, format conversion or OCR, by directly using document screenshots. Representative methods include DSE~\cite{ma2024unifying}, a single embedding bi-encoder model, and ColPali~\cite{faysse2024colpali}, a ColBERT-like~\cite{khattab2020colbert} multi-vector embedding dense retriever. These vision-based dense retrievers can achieve retrieval effectiveness comparable to strong text-based dense retrievers in text-intensive retrieval settings and demonstrate superior effectiveness in vision-based retrieval scenarios, such as slide retrieval. We note that these document screenshot retrievers are different from multi-modal retrievers such as CLIP~\cite{radford2021learningtransferablevisualmodels}, which are designed to handle natural images rather than text-intensive document screenshots.

VLM-based document screenshot retrievers are increasingly being integrated into retrieval augmented generation pipelines~\cite{yu2024visrag,ma2024visa,riedler2024beyond,xia2024mmed,cho2024m3docrag}. %, and are being augmented with additional capabilities such as attribution components~\cite{bibid}.
 In this paper, we empirically demonstrate that DSE and ColPali vision-based dense retrievers are vulnerable to attacks that can be exploited for corpus poisoning or SEO purposes.

\subsection{Corpus Poisoning and SEO Attacks}
Corpus poisoning refers to the deliberate manipulation or alteration of a dataset (corpus)~\cite{wang2022threats,tian2022comprehensive,das2024security,liu2024robust}; the goal is often to influence the performance, behaviour, or output of models operating on the poisoned corpus in a way that benefits an attacker or serves a specific agenda. In information retrieval, this consists of injecting manipulated documents into a corpus to skew rankings or scoring. While in other areas of Artificial Intelligence corpus poisoning is predominantly performed to manipulate training data, in IR this is often performed to affect the corpus on which retrieval takes place~\cite{liu2024robust}; e.g., so that poisoned documents ranks highly in search engines, skewing the results toward irrelevant or biased content. Corpus poisoning ultimately impacts user experience and erodes trust in the search engine. Next we describe an example of corpus poisoning attack specifically for dense retrievers.

The first corpus poisoning attack for dense retrievers was a text perturbation approach~\cite{zhong-etal-2023-poisoning}, inspired by the HotFlip method~\cite{ebrahimi2018hotflip}. This method generates a small set of adversarial passages by perturbing discrete tokens in randomly initialised passages to maximise their similarity with a provided set of training queries. These adversarial passages are then inserted into the retrieval corpus, and the success of the attack is determined by the retrieval of these adversarial passages at top rank positions in response to future unseen queries. %These adversarial passages can be used to harm retrieval effectiveness and/or inject spam or misinformation into the search engine result list.

The HotFlip attack required access to the dense retriever's model weights. A more dangerous model that did not require model weights is BaD-DPR, which uses grammar errors as triggers for retrieval of poisoned documents~\cite{long2024backdoor}.

Vec2Text is a method to invert text embeddings, recovering the original text from an emdedding~\cite{morris-etal-2023-text,zhuangv2t}. Vec2Text has also been proposed as a corpus poising attach method~\cite{zhuang2024doesvec2textposenew}. This method first trains a Vec2Text model that is able to reconstruct text from a given embedding. Then, given training queries, the embeddings closest to all the query embeddings are computed and sent to the Vec2Text model for adversarial document generation. Similar to BaD-DPR, this method does not require access to the dense retriever's weights.

%All the existing corpus poisoning methods designed for traditional text-based dense retrievers. 
In this paper, we are the first to explore attacks on vision-based dense retrievers, as opposed to text-based dense retrievers. Unlike text-based attacks, where gradient information cannot be directly propagated to the text input due to the vocabulary lookup step in tokenisation, vision-based dense retrievers operate on document representations as screenshot pixels. This allows gradients to be directly applied to the pixels, enabling the manipulation of the document.

The practice of Search Engine Optimisation (SEO) shares similarities with corpus poisoning, though the two often differ in intent. While corpus poisoning is adversarial, aiming to undermine the integrity of retrieval systems, SEO focuses on optimising specific content to achieve better rankings in search engines. However, overlaps arise when unethical ``black hat'' SEO techniques exploit system vulnerabilities to gain an unfair advantage~\cite{castillo2011adversarial} -- this is the scenario targeted by our attacks. The key distinction lies in the attacker’s objectives: in SEO, the attacker aims to ensure that users engage with their content, avoiding methods that render the retrieved document unreadable or obviously compromised. In contrast, corpus poisoning generally does not prioritise the usability of the retrieved document, unless the retrieval system employs defence mechanisms that explicitly address this aspect.

%\subsection{Adversarial images in computer vision}

\section{Method}

We design our attack methods by considering the potential objectives of a malicious attacker. In this context, an attacker typically has two primary goals: ensuring the injected document ranks highly in the search engine (attack \textit{effectiveness}) and maintaining the manipulated document’s similarity to the original content (attack \textit{fidelity}). Attack effectiveness is crucial for both corpus poisoning and SEO tasks. In corpus poisoning, the goal is to disrupt the ranker’s performance, while in SEO, the goal is to promote the attacker’s document. Attack fidelity is particularly important in the SEO context, as the attacker needs the document to remain readable and avoid detection by users, so that the trust in the content of the document is not affected. In the case of corpus poisoning, fidelity is generally less critical since the primary goal is to degrade system effectiveness. However, maintaining fidelity can still be advantageous for evading defence mechanisms implemented by the search engine.

Intuitively, there is a trade-off between effectiveness and fidelity. A method that allows greater freedom to alter the content (i.e., lower fidelity) is likely to achieve higher attack effectiveness. Conversely, maintaining higher fidelity often limits the potential for attack effectiveness, as it restricts the extent of permissible changes. Our proposed attack methods include a parameter that allows for a flexible balance between these two objectives.

In devising the attack methods, we assume the attacker has access to the retriever’s weights at inference but cannot modify the model itself. Given this constraint, we explore white-box gradient-based attacks, which leverage gradient information to optimise an adversarial document screenshot to rank highly in the search results. Gradient-based methods have been used in adversarial attacks on vision models, demonstrating their effectiveness in manipulating model outputs while controlling perturbation magnitude.

As a base, we adopt the Fast Gradient Sign Method (FGSM)~\cite{Goodfellow2015fgsm}, a widely used gradient-based attack that perturbs inputs by adjusting them in the direction of the gradient’s sign with respect to the loss function. Various extensions of FGSM have been developed to enhance its effectiveness~\cite{Kurakin2016ifgsm,Dong2017mifgsm,Lin2020Nesterov,Xiaosen2021variance}. The Iterative Fast Gradient Sign Method (I-FGSM)~\cite{Kurakin2016ifgsm} refines FGSM by applying multiple smaller updates, producing stronger adversarial perturbations; we exploit a similar mechanism in our approaches, where we also iterate using small steps. The Momentum Iterative Fast Gradient Sign Method (MI-FGSM)~\cite{Dong2017mifgsm} introduces a momentum term to stabilise gradient updates, improving attack transferability across models and avoiding poor local optima; in our approaches we exploit a similar mechanism of gradient normalisation. %The Scale-Invariant Nesterov Iterative Fast Gradient Sign Method (SI-NI-FGSM)~\cite{Lin2020Nesterov} incorporates Nesterov’s accelerated gradient technique and a scale-invariant property to enhance attack effectiveness across different model scales. Additionally, the Variance Tuning Method~\cite{Xiaosen2021variance} modifies gradient updates by introducing variance tuning, further improving transferability by preventing overfitting to specific decision boundaries.

Building on these approaches, we adapt FGSM to our task and propose three variations of the attack, each designed to balance effectiveness and fidelity differently: direct optimisation, noise optimisation, and mask optimisation. We detail each approach below, describing its core mechanism and outlining strategies to enhance fidelity -- though these often come at the cost of reduced attack effectiveness.

\subsection{Direct Optimisation}

The Direct Optimisation method follows a process similar to FGSM, iteratively modifying the pixel values of an adversarial document screenshot by adjusting them in the direction of the gradient’s sign to maximise ranking effectiveness.
Formally, let $x \in \mathbb{R}^{H, W, C}$  represent an adversarial document screenshot that we aim to rank highly in the search results; where $H, W$ and $C$ denote the dimensions of the original screenshot. Let $C$ be the target corpus, $\mathcal{R}$ the target retriever; $\mathcal{L}$ the corresponding loss function.
The optimisation process starts with the initial image $x_0 = x$.

At each iteration, the document is updated using the gradient of the loss function as follows:

$$x_{i+1} = Clip \left[ x_i - \alpha \cdot sign( \dfrac{\nabla_x \mathcal{L}(x_i, C) }{||\nabla_x  \mathcal{L}(x_i, C) ||}) \right]$$

Here, the $Clip(\cdot)$ function constrains pixel values within valid bounds, ensuring that the generated adversarial example remains a realistic document screenshot, and the $sign$ function determines the direction of gradient updates.
% $\alpha$=1 and follows a cosine annealing schedule.

In order to minimise the effect of the optimisation process on the visual appearance of the image, we propose to use only the top-$p$ percentage of the gradient at each step:

$$\widetilde{\nabla}_x  = \nabla_x \odot 1_{Top-p(|\nabla_x|)}  $$

This approach reduces visual distortions by limiting modifications to the most influential pixels at each step, ensuring that changes are spread across different regions of the image rather than concentrated in a single area (see Figure~\ref{direct_optimization} as an example). As a result, although the attack remains noticeable, the readability of the document improves, while maintaining a high level of attack effectiveness.

%\subsubsection{Whole-of-Image Pixels Optimization}
%
%\subsubsection{Localized Pixels Optimization}

\subsection{Noise Optimisation}

Unlike Direct Optimisation, which modifies the document pixels directly, the Noise Optimisation method learns an additive noise pattern that, when applied to the image, increases its ranking while preserving fidelity. Instead of altering the image itself, we optimise a noise image that is iteratively refined to maximise the attack's effectiveness.

We initialise the noise image as a zero matrix: $n  \in \mathbb{R}^{H, W, C}, n=0 $.
At step 0: $n_0 = n$. Then, at each iteration, the noise is updated as follows:

$$n_{i+1} = Clip\left[ n_i + \alpha \cdot sign( \dfrac{\nabla_n \mathcal{L}(x + n_i, C) }{||\nabla_n  \mathcal{L}(x + n_i, C) ||}) \right]$$

\noindent The final adversarial document screenshot is then computed as $x + n$.

As in the Direct Optimisation method, we aim to minimise the perceptibility of the attack by modifying only the most influential pixels at each step. However, in this case, the top-$p$ gradient selection is applied to the noise image rather than the original document: $\widetilde{\nabla}_n  = \nabla_n \odot 1_{Top-p(| \nabla_n  |)}  $

Visually, this attack is less noticeable than Direct Optimisation (Figure~\ref{noise_optimization}). Instead of introducing concentrated distortions, it produces a blurry appearance, resembling poor image quality due to buffering or incomplete page loading. However, as shown in our experiments, this method is less effective than Direct Optimisation in ranking manipulation.

\subsection{Mask Direct Optimisation}

The Mask Direct Optimisation method uses similar optimisation algorithm as Direct Optimisation. However, instead of using the original screenshot image, we first resize it to a smaller size and then add a white mask margin around it to restore the image to its original dimensions. During the optimisation process, only the pixels within this mask margin are updated, leaving the original image completely untouched and preserving all its information.

Formally, let $x_s \in \mathbb{R}^{h,w,C}$ be a resized version of the original adversarial screenshot $x \in \mathbb{R}^{H,W,C}$, where $h=aH$ and $w=aW$ for some mask area percentage $a \in [0\%, 100\%]$. The mask area percentage defines the size of mask margin in relation to the size of the original screenshot document. We define a binary mask $M \in \{0,1\}^{H\times W}$, where $M(i,j)=1$ inside the mask region of size $(H-h) \times  (W-w)$, and $M(i,j)=0$ elsewhere. Let $P(x_s)$ denote the resized image placed inside the mask.

The modified image is:

$$x_0 = x \odot (1 - M) + P(x_s) \odot M$$

This modified image is used for optimisation as follows:

\begin{equation}
	x_{i+1} = Clip \left[ x_i - \alpha \cdot sign( \dfrac{\widetilde{\nabla}_x\mathcal{L}(x_i, Q) }{||\widetilde{\nabla}_x \mathcal{L}(x_i, Q) ||}) \right]
	\label{eq:mask_optimization}
\end{equation}

where only the mask gradient is updated:

$$\widetilde{\nabla}_x  = \nabla_x \odot 1_{(1-M)}  $$

Note that the mask $M$ does not depend on the gradient $\nabla_x$; it is pre-defined and remains static throughout the optimisation.

This approach ensures that the original content remains intact, side from being of reduced size and surrounded by the mask. We note that this type of mask-based manipulation is uncommon in the computer vision field -- most adversarial attack methods directly modify the image pixels. However, this approach aligns better with the information retrieval setting, as it ensures that all the original image information remains unchanged and visible to the user.

\begin{figure*}[]
	\centering
	\includegraphics[width=0.32\textwidth]{images/original_image/original_img.png}
	\includegraphics[width=0.32\textwidth]{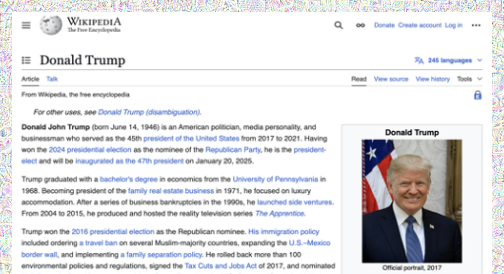}
	\includegraphics[width=0.32\textwidth]{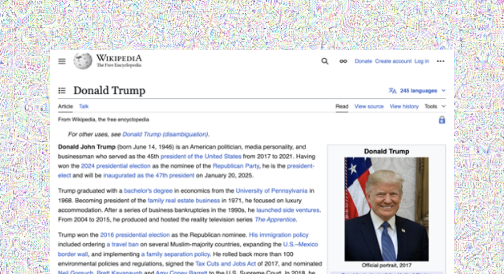}

	\caption{
		Mask Optimisation. Left: original image, middle: 5\% mask; right: 20\% mask.}
	\label{mask_optimization}
\end{figure*}

To balance the trade-off between attack fidelity and effectiveness, we make the mask margin area $a$ adjustable, scaling it proportionally to the original image’s width and height. Our experiments primarily focus on values of $a$ ranging from 5\% of the original image (where the original image occupies most of the space) to 100\% (where the mask replaces the entire image). This adjustment provides precise control over the number of pixels that can be modified during the optimisation process. Figure~\ref{mask_optimization} illustrates examples generated with different mask sizes.

%\todo{is this method used in CV?}

\section{Experimental Setup}

\begin{figure*}[]
	\centering
	\includegraphics[width=0.32\textwidth]{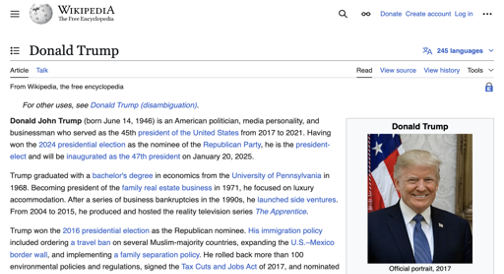}
	\includegraphics[width=0.32\textwidth]{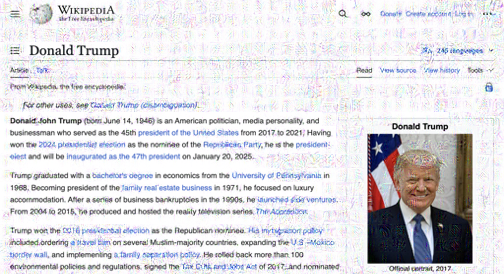}
	\includegraphics[width=0.32\textwidth]{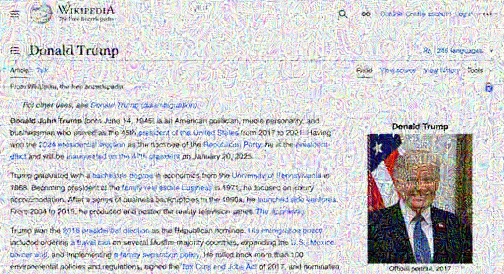}
	\caption{
		Direct Optimisation. Left: original image, middle: 10\% gradient is updated; right: 100\% gradient is updated. }
	\label{direct_optimization}
\end{figure*}

\begin{figure*}[]
	\centering
	\includegraphics[width=0.32\textwidth]{images/original_image/original_img.png}
	\includegraphics[width=0.32\textwidth]{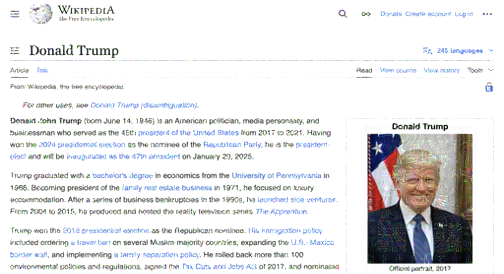}
	\includegraphics[width=0.32\textwidth]{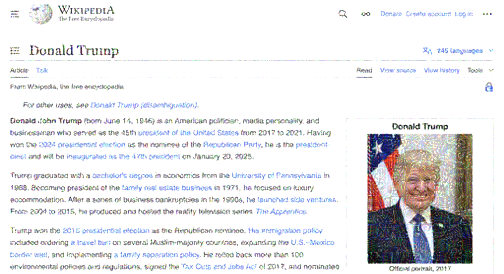}
	\caption{
		Noise Optimisation. Left: original image, middle: 10\% gradient is updated; right: 100\% gradient is updated.}
	\label{noise_optimization}
\end{figure*}

\subsection{Attack Settings} \label{sec_attack_setup}
To investigate the behaviour of VLM-based document retrievers, we consider experimental settings of increasing attack difficulty: from targetting a small set of known queries, to unknown queries from the same distribution of the queries used for training, to unknown out-of-domain queries and documents. These settings closely mirror a range of real-world scenarios. %, which we further discuss in Section~\ref{sec_discussion}. 

%In this section, we outline the attack settings explored in our study. These settings are designed to simulate real-world scenarios, progressing from easier to more challenging conditions. The objective is to thoroughly evaluate the method’s performance in achieving a balance between attack effectiveness and fidelity.

\subsubsection{Attacking Predefined Target Queries} \label{sec_target_query_setup}
%In this setting, we assume that the attacker has a group of target queries for which they want to optimize the ranking of their own documents, which reflects a typical SEO scenario. For example, during an election campaign,  a supporter of Donald Trump might aim to increase the visibility of his Wikipedia page while undermining his competitor Joe Biden. The attacker’s goal could be to ensure that when users search for queries related to Joe Biden, Trump’s page appears as the top-ranked document. In this case, the targeted query group consists of Joe Biden-related search queries, and the optimized document screenshot is Trump’s Wikipedia page.

We start by considering the easiest attack setting, where the attacker is targeting a focused, predefined group of target queries. To operationalise this, we randomly selected 10 queries from the Wiki-SS-NQ test query set (see Section~\ref{sec_dataset} for a description of this dataset). Each selected query, along with its corresponding answer, was provided to ChatGPT to generate 9 additional similar queries. This process resulted in 10 groups, each  consisting of 10 target queries focused around a similar topic.

Then, we considered the choice of documents that the methods should manipulate to carry on the attack -- we refer to these as seed documents. To sample seed documents, we executed all the test queries from the Wiki-SS dataset against the DSE retriever; we then fused the individual rankings into a unique ranking using reciprocal rank fusion.
From the fused ranking, we selected the bottom 100 documents (i.e. those at rank $\approx$1.2 million): these are on average the most irrelevant documents to the test queries in Wiki-SS-NQ. We selected these documents because they represent the most challenging setting where the attack must optimise highly irrelevant documents for the queries in the Wiki-SS dataset. Finally, we evaluated our attack methods by independently manipulating (i.e. optimising with an attack method) one of the 100 sampled documents with respect to a target query group, injecting then the manipulated document into the dataset for retrieval, and measuring where the manipulated document is retrieved in answer to the specific target queries considered. We executed this then for every query group and seed document, and averaged their results.

This setting resembles a common (black-hat) SEO scenario, although it can also be considered across broader corpus poisoning scenarios. An example of a typical SEO scenario is as follows: A seller on an e-commerce platform might attempt to increase the visibility of their new smartphone model while reducing the prominence of competitors' products. The attacker's goal could be to ensure that when users search for queries like "best budget smartphone," their smartphone model appears as the top-ranked product. In this case, the targeted query group consists of budget smartphone-related search queries, and the optimized product page is the attacker's smartphone listing.

\subsubsection{Attacking In-Domain Queries} \label{sec_indomain_query_setup}
Next, we consider a harder attack setting where instead of a restricted set of known queries, the goal is to attack a large number of queries, from a known distribution (but the queries themselves are not known a priori by the attacker). We operationalised this setup using the queries distributed with the Wiki-SS-NQ dataset, relying on the training portion to optimise the seed documents according to our attacks, and on the test portion for validating the effectiveness of the attack. The test queries are disjoint from the training queries; however they are sampled from the same distribution of queries, and thus can be considered as in-domain. 
As seed documents to manipulate during the attach, we used the 100 documents sampled using the methodology described for the previous attack. Manipulated documents were injected in the ranking, and their position in the retrieval list across all test queries was recorded to compute attack effectiveness.

This setting parallels a typical corpus poisoning scenario, where the attacker aims to degrade the search engine’s performance across all queries, and in fact aligns with the corpus poisoning task described in the literature~\cite{zhong-etal-2023-poisoning, zhuang2024doesvec2textposenew, li2025reproducing}, where an attacker injects adversarial documents into the index to manipulate the search engine. The assumption is that the attacker has access to a query log representative of the queries commonly submitted to the search engine. This setting can also resemble a far-reaching SEO tactic in which a document is promoted for any query; however, in practice, SEO efforts generally focus on optimising for specific, high-value queries.

\subsubsection{Attacking an Out-Of-Domain Dataset} \label{sec_outofdomain_query_setup}
Finally, we examine the most challenging of the three experimental settings in our study. Here, the goal is to attack queries and documents from an unknown distribution. In other words, not only might the test queries differ substantially from the distribution of the training queries, but the documents being retrieved may also vary greatly from the single document that has been manipulated by the attack and injected into the retrieval corpus. We operationalised this scenario using the Vidore benchmark in a zero-shot setup (see Section~\ref{sec_dataset} for details of this benchmark). Specifically, we used Wiki-SS-NQ training queries to optimise the attack on a Wiki-SS-NQ document, then inserted the manipulated document into the Vidore dataset and ran the corresponding Vidore queries to measure the effectiveness of the attack.

This setting mirrors a challenging corpus poisoning scenario, in which the attacker seeks to undermine the search engine’s performance across all queries while lacking knowledge of the documents it indexes and the queries it typically handles.

%\subsubsection{Corpus poisoning attack}
%
%\todo{We use Wiki-cc dataset, which has xx number of training queries, and xx number of testing queries. The corpus is 1m+ wikipeida webpage screenshot.}
%
%\todo{The task is use training queries of Wiki-cc to train the adversarial images. Add the image to the corpus and then do retrieval testing queries and see if the adversarial images could be retrieved on the top. Evaluation is top-k success.}
%
%\subsubsection{Zero-shot Generalisation}
%\todo{Zero-shot generalization: add adversarial images generated with wiki-cc to Vidreo benckmark to test if Vidreo datasets could be attacked.}

\vspace{-6pt}
\subsection{Datasets and Evaluation}
\label{sec_dataset}
Our experiments use the Wiki-SS-NQ~\cite{ma2024unifying} and the Vidore benchmarks~\cite{faysse2024colpali}. Both are benchmarks for the document screenshot retrieval task.

Wiki-SS-NQ is based on Google's Natural Questions dataset~\cite{kwiatkowski-etal-2019-natural}, where the original training and testing queries are retained. However, the original textual documents, which are the contents of Wikipedia pages, have been replaced with corresponding Wikipedia webpage screenshots. The dataset consists of around 30k training queries and 3,610 test queries, with a corpus of approximately 1.2 million document screenshots. We use this dataset when studying attacks to predefined target queries and in-domain queries.

%sampling target queries and evaluating attack effectiveness in an in-domain setting.
The Vidore benchmark is a collection of 10 screenshot retrieval datasets. It includes multiple domains (medical, business, scientific, administrative), languages (English and French), and modalities (text, figures, infographics, tables) along with both real and synthetically generated queries. We use this dataset when evaluating attack effectiveness in an out-of-domain setting. %Specifically, we train the adversarial screenshots using Wiki-SS-NQ training queries, inject the adversarial screenshots into the Vidore datasets, and then assess their attack effectiveness.

To evaluate attack effectiveness, we follow previous corpus poisoning literature~\cite{zhong-etal-2023-poisoning, zhuang2024doesvec2textposenew} and use top-k attack success rate (success@k): the proportion of queries for which at least one adversarial passage is retrieved in the top-k results. A higher success rate indicates that the model is more vulnerable to attacks. We consider shallow values of k (k=1, 5, 10) because these represent different portions of a typical search engine result page that are commonly examined by users. We also consider k=100, a ranking cut-off that is more commonly considered when these top retrieved documents are used within a broader pipeline (e.g., for re-ranking, or long-context retrieval augmented generation).
In addition, we also report the mean reciprocal rank of attack at rank 100 (MRRA@100): for this we only consider the top 100 ranked documents, find the first occurrence of the manipulated documents in the ranking (we consider setting both settings with a single and multiple manipulated documents), and record its rank -- then MRRA@100 is the average of the reciprocal rank positions that have been recorded across queries. 
 %treating the adversarial screenshots as the only relevant documents for the targeted queries. 
This evaluation reflects how highly the search engine ranks the adversarial screenshots within the results.

\subsection{Implementation Details}
We conduct the attack experiments on two popular document screenshot retrievers: DSE~\cite{ma2024unifying}\footnote{\url{https://huggingface.co/MrLight/dse-qwen2-2b-mrl-v1}}, a single-embedding bi-encoder model, and ColPali~\cite{faysse2024colpali}\footnote{\url{https://huggingface.co/vidore/colpali-v1.2-hf}.}, a ColBERT-like~\cite{khattab2020colbert} multi-vector embedding dense retriever. Both model checkpoints are publicly available on the Hugging Face model hub. We use the Tevatron~\cite{tevatron} information retrieval toolkit for encoding and retrieval on the tested datasets. Each training session for adversarial images is conducted on a single H100 GPU with FlashAttention~\cite{flashattention} to optimise GPU memory usage. Due to the large number of experimental settings, we distribute each training job across multiple GPUs.

For all the training, we use standard gradient descent to calculate the gradient on pixels and train on the entire set of training queries. The retrieval model parameters are frozen during the training. We employ cosine learning rate decay, starting with a learning rate of 1, and train for 3,000 steps. Training a single adversarial image takes approximately 10 minutes for DSE and 20 minutes for ColPali.

\section{Results}
\label{sec_results}

\subsection{Results of Attacking Predefined Target Queries} \label{sec_results_target_queries}
We start by discussing the results obtained in the context of targeting a predefined group of queries. Recall that in this setting, we randomly sampled 10 queries from the Wiki-SS-NQ test query set and generated 9 additional similar queries for each sampled query, resulting in 10 groups of target queries. We then optimised each of the 100 most irrelevant document screenshots from the Wiki-SS-NQ dataset, with the goal to rank these documents at the top for the 10 groups of queries individually, reporting the average performance across query groups and screenshots (see Section~\ref{sec_target_query_setup}).

Table~\ref{tab:ablation_seed_documents_first_experiment} presents the results under a configuration where the optimised gradient $p$ is set to 100\% for direct and noise optimisation. This configuration allows the methods to focus solely on attack effectiveness without being constrained by fidelity, providing insight into the upper bound of attack performance for our methods. For the mask direct optimisation attack, results in the table are obtained for a mask area $a$ of 20\% the size of the original screenshot: this can be considered already a large mask for this type of attack, as it is easily visible for the user. %This setup also mimics a real-world scenario where an attacker aims to poison a group of queries without prioritizing fidelity.

Notably, in this setting all attack methods achieve complete success across all queries and for all seed documents (success@1=1.0). We stress that the standard deviation across all attack attempts is 0 (10 query groups × 3 attack methods × 100 screenshots = 3,000 attempts) indicating that every attack attempt successfully generated an adversarial image ranked at the top position. These results highlight the significant vulnerability of document screenshot retrievers to our proposed pixel-level attacks when the attacks are optimised for a small, known set of target queries.

%As demonstrated by the table, the success@1 rate is 100\% across all our methods and models. Notably, the standard deviation is consistently 0, indicating that every attack attempt (10 query groups × 3 attack methods × 100 screenshots = 3,000 attempts) successfully generated an adversarial image ranked at the top position. These results highlight the significant vulnerability of document screenshot retrievers to our proposed pixel-level attacks.

\begin{table}[t]
	\centering
		\caption{Results obtained when attacking predefined target queries. We report the mean and standard deviation for success@1 across 10 target query groups and 100 irrelevant document screenshots.}
	\begin{tabular}{l|c|c}
		\toprule
		Method & DSE & Colpali\\
		\midrule
		Direct ($p=100\%$) & 1.0 $_{\pm0.0}$ & 1.0 $_{\pm0.0}$\\
		Noise ($p=100\%$) & 1.0 $_{\pm0.0}$ & 1.0 $_{\pm0.0}$\\ 
		Mask ($a=20\%$) & 1.0 $_{\pm0.0}$& 1.0 $_{\pm0.0}$\\
		\bottomrule
	\end{tabular}
	\label{tab:ablation_seed_documents_first_experiment}
\end{table}

\begin{figure}[t]
	\centering
	\includegraphics[width=0.49\textwidth]{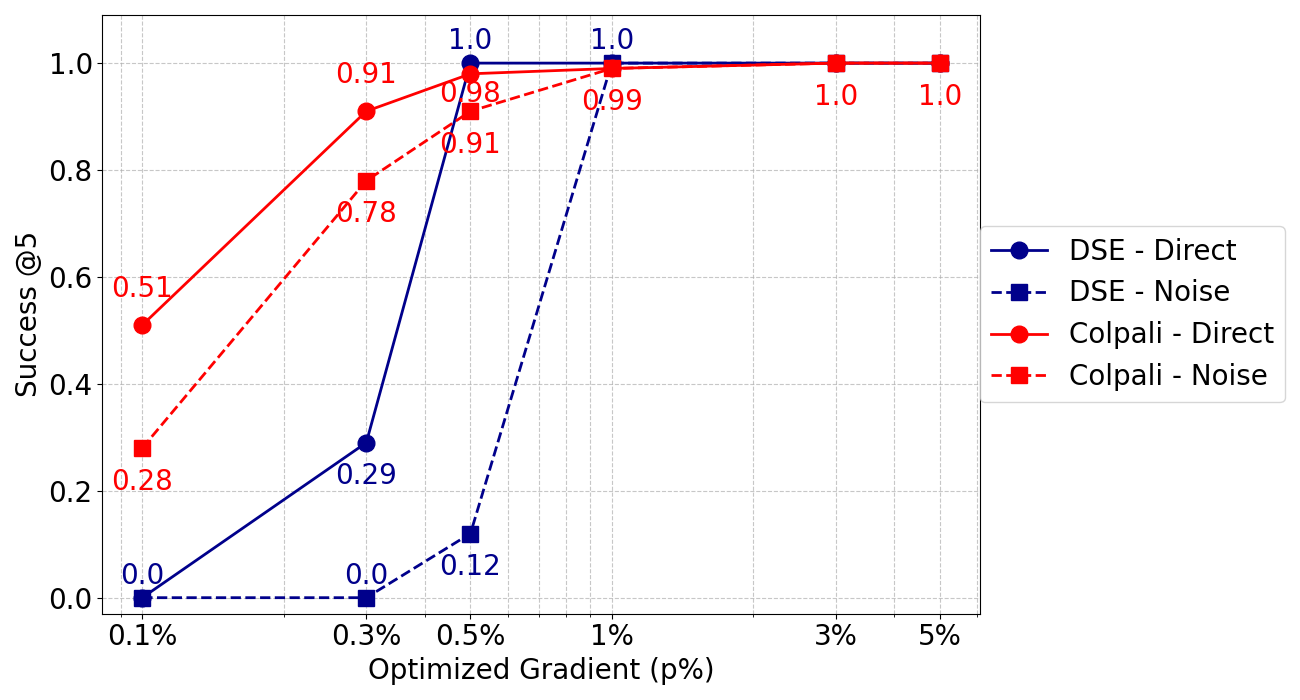}
	%\resizebox{\textwidth}{!}{%
		%		\begin{tabular}{c|c}
			%			\includegraphics[width=0.45\textwidth]{images/ablation_grad_perc_k1_dse.png}& \includegraphics[width=0.45\textwidth]{images/ablation_grad_perc_k1_colpali.png}
			%		\end{tabular}
		%		%}
	\caption{Impact of gradient optimisation percentage $p$ on attack effectiveness over target queries. Lower percentages of optimised gradient $p$ result in less visual corruption of the document screenshot.\vspace{-10pt}	}
	\label{Ablation:grad_percentage_exp1}
\end{figure}
\begin{figure}[t]
	\centering
	\includegraphics[width=0.49\textwidth]{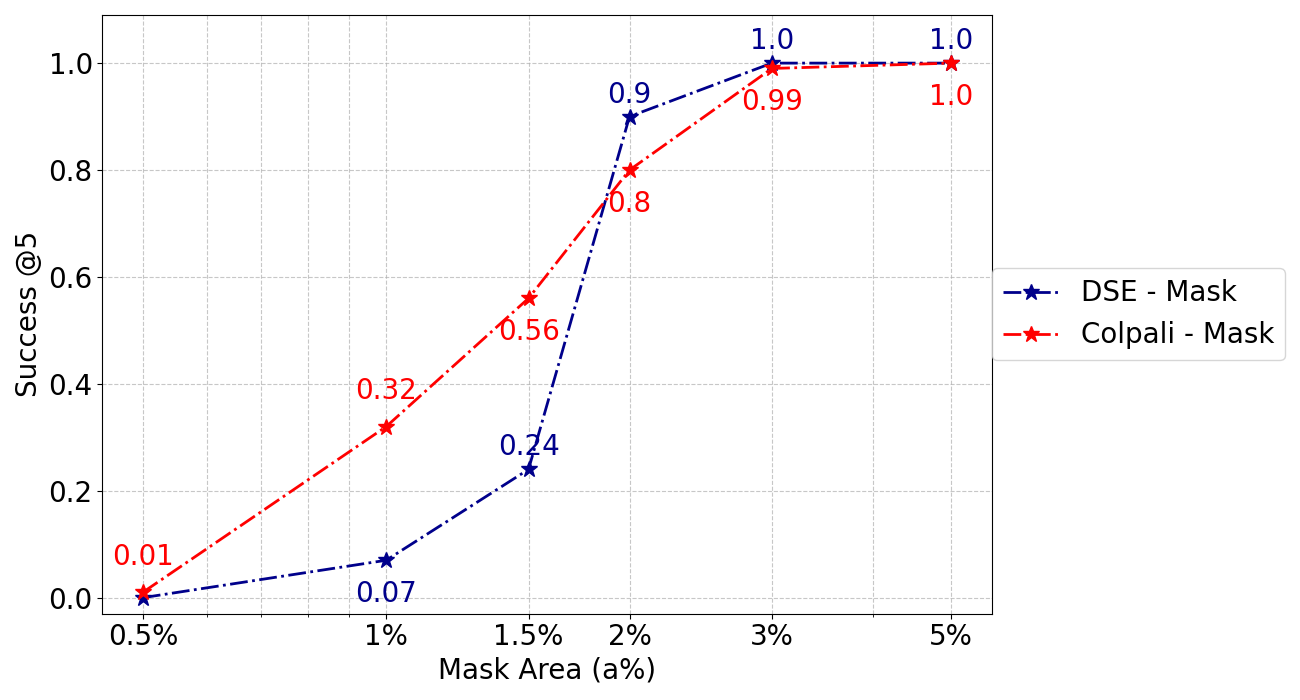}
	
	\caption{Impact of mask area $a$ on attack effectiveness over target queries. Lower percentages of mask area $a$ result in less visual corruption of the document screenshot. \vspace{-10pt}}
	\label{Ablation:mask_percentage_exp1}
\end{figure}

\begin{table*}[t!]
	\centering
		\caption{Results obtained when attacking a large set of in-domain queries. We report the mean and standard deviation for success@k and MRRA@100 across 10 target query groups and 100 irrelevant document screenshots.}\vspace{-5pt}
	\resizebox{0.9\textwidth}{!}{%
		{\renewcommand{\arraystretch}{1.2}
			\begin{tabular}{l|l|ccccc}
				\toprule
				Model&Method & \textbf{s@1} & \textbf{s@5} & \textbf{s@10} & \textbf{s@100}  & \textbf{MRRA@100}  \\\midrule
				%				& \multicolumn{5}{c}{\textbf{DSE}}  \\
				% & \multicolumn{12}{c}{\textbf{k=1}} \\
				\multirow{3}{*}{\textbf{DSE}}&Direct ($p = 100\%$) & 0.1177 $\pm$ 0.0010 & 0.3241 $\pm$ 0.0009 & 0.4186 $\pm$ 0.0009 & 0.6999 $\pm$ 0.0011 & 0.2154 $\pm$ 0.0007 \\
				%  & 0.3319 & 0.4246 & 0.515 & 0.7047 &  &  &  &  \\
				&Noise ($p = 100\%$)& 0.1144 $\pm$ 0.0031 & 0.3185 $\pm$ 0.0041 & 0.4097 $\pm$ 0.0047 & 0.6904 $\pm$ 0.0053 & 0.2110 $\pm$ 0.0033  \\
				&Mask ($a = 20\%$)& 
				0.1082 $\pm$ 0.0045 & 0.3093 $\pm$ 0.0065 & 0.4003 $\pm$ 0.0072 & 0.6802 $\pm$ 0.0072 & 0.2038 $\pm$ 0.0051 			\\\midrule
				
				%				& \multicolumn{5}{c}{\textbf{Colpali}}\\
				
				\multirow{3}{*}{\textbf{ColPali}}&Direct ($p = 100\%$)& 0.1028 $\pm$ 0.0065 & 0.2045 $\pm$ 0.0100 & 0.2640 $\pm$ 0.0111 & 0.5203 $\pm$ 0.0113 & 0.1608 $\pm$ 0.0076 \\
				&Noise ($p = 100\%$)& 0.0213 $\pm$ 0.0065 & 0.0590 $\pm$ 0.0141 & 0.0892 $\pm$ 0.0182 & 0.2785 $\pm$ 0.0370 & 0.0515 $\pm$ 0.0100  \\
				&Mask ($a = 20\%$) & 0.0274 $\pm$ 0.0084 & 0.0683 $\pm$ 0.0166 & 0.0988 $\pm$ 0.0200 & 0.2833 $\pm$ 0.0335 & 0.0588 $\pm$ 0.0119 \\

				\bottomrule
			\end{tabular}
		}
	}
	\label{tab:results_nq}\vspace{-5pt}
\end{table*}

We further conducted ablation studies to examine the impact of parameters that control attack fidelity on the effectiveness of the attacks. For this setting, we explored small values for the optimised gradient $p$, ranging from 0.1\% to 5\% for direct and noise optimisation, and from 0.5\% to 5\% for the direct optimisation of the mask area $a$~\footnote{The smallest mask value in our experiments is 0.5\%, which corresponds to a single-pixel-wide margin around the image.}. These parameter ranges allow for achieving high attack fidelity (i.e. making the attack barely visually noticeable by the user).

%Due to the large number of experiments required, 

For our ablation studies, we optimised only a single seed document screenshot\footnote{Donald Trump's Wikipedia page, as shown in Figures~\ref{direct_optimization}-\ref{mask_optimization}.} for the 10 target query groups. We did this because studying multiple seed documents would be computationally intractable in many of the cases considered by the ablation studies. However, we expect the results obtain on a single seed document to generalise to other samples: this is because we observed results reported in Table~\ref{tab:ablation_seed_documents_first_experiment} displayed no variance across different seed documents. 
%The results are expected to generalize to other images, as demonstrated in Table~\ref{tab:ablation_seed_documents_first_experiment}, where the standard deviation was low, even for the most irrelevant documents. 

The results of the ablation studies are reported in Figures~\ref{Ablation:grad_percentage_exp1} and \ref{Ablation:mask_percentage_exp1}. For both DSE and ColPali retrievers, high attack effectiveness (complete success@5) can be achieved with very small values of the optimised gradient (0.5\% to 1\%) for direct and noise optimisation methods, and with a mask area of 3\% for mask direct optimisation (though already $2\%$ displays strong attack effectiveness). These findings suggest that if an attacker is targeting a set of predefined queries, high attack effectiveness can be achieved while maintaining high attack fidelity -- and thus stealthiness.
Additionally, ColPali appears to be more vulnerable than DSE in this setting, as it requires smaller values of the optimised gradient and mask area to achieve higher success@5 rates compared to DSE.

\subsection{Results of Attacking In-Domain Queries} \label{sec_results_indomain_queries}

Next, we consider the results obtained when attacking the VLM-based document retrievers across a large set of in-domain queries. For this, we trained the document screenshots to optimise their ranking across all Wiki-SS-NQ training queries. We then evaluated the optimised documents on in-domain unseen Wiki-SS-NQ test queries. %This setting aligns with the corpus poisoning task described in the literature~\cite{zhong-etal-2023-poisoning, zhuang2024doesvec2textposenew, li2025reproducing}, where an attacker injects adversarial documents into the index to manipulate the search engine. %In such scenarios, the goal of an attacker is that, regardless of the user's search query, the injected adversarial document is retrieved, thereby degrading the user experience.

In this setting, attack fidelity (the similarity between the attacked document and the original) is often not a priority. Therefore, we use a $p=100\%$ for optimising the gradient for the direct and noise optimisation methods and a mask area of $a=20\%$ for the mask direct optimisation method. We again optimise the 100 most irrelevant document screenshots with respect to the Wiki-SS-NQ test queries and report the average performance across these screenshots.

The results are reported in Table~\ref{tab:results_nq}. Even injecting a single adversarial document screenshot enables all the investigated attack methods to effectively compromise both VLM-based retrievers. For example, our direct optimisation method successfully generated adversarial screenshots ranked in the top-10 for 41.86\% of unseen queries in the case of DSE. For ColPali, this success rate drops to 26.4\%, but it is still substantial (affecting about 1 in 4 queries). The low standard deviation values in the results indicate that the choice of the seed document screenshot has minimal impact on the attack's effectiveness.

\begin{figure}[t]
	\centering
	\includegraphics[width=0.49\textwidth]{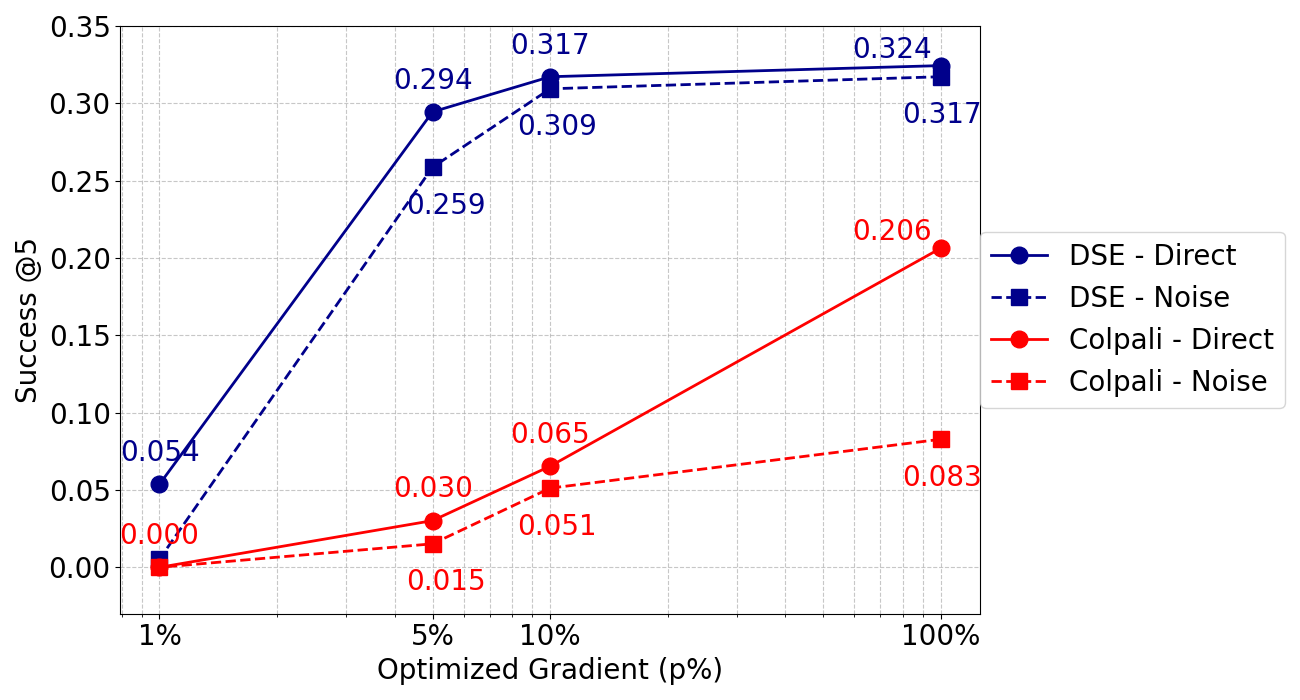}
	%\resizebox{\textwidth}{!}{%
		%		\begin{tabular}{c|c}
			%			\includegraphics[width=0.45\textwidth]{images/ablation_grad_perc_k1_dse.png}& \includegraphics[width=0.45\textwidth]{images/ablation_grad_perc_k1_colpali.png}
			%		\end{tabular}
		%		%}
	\caption{Impact of gradient optimisation percentage $p$ on attack effectiveness over in-domain queries. Lower percentages of optimised gradient $p$ result in less visual corruption of the the document screenshot. \vspace{-10pt}}
	\label{Ablation:grad_percentage}
\end{figure}

\begin{figure}[t]
	\centering
	\includegraphics[width=0.49\textwidth]{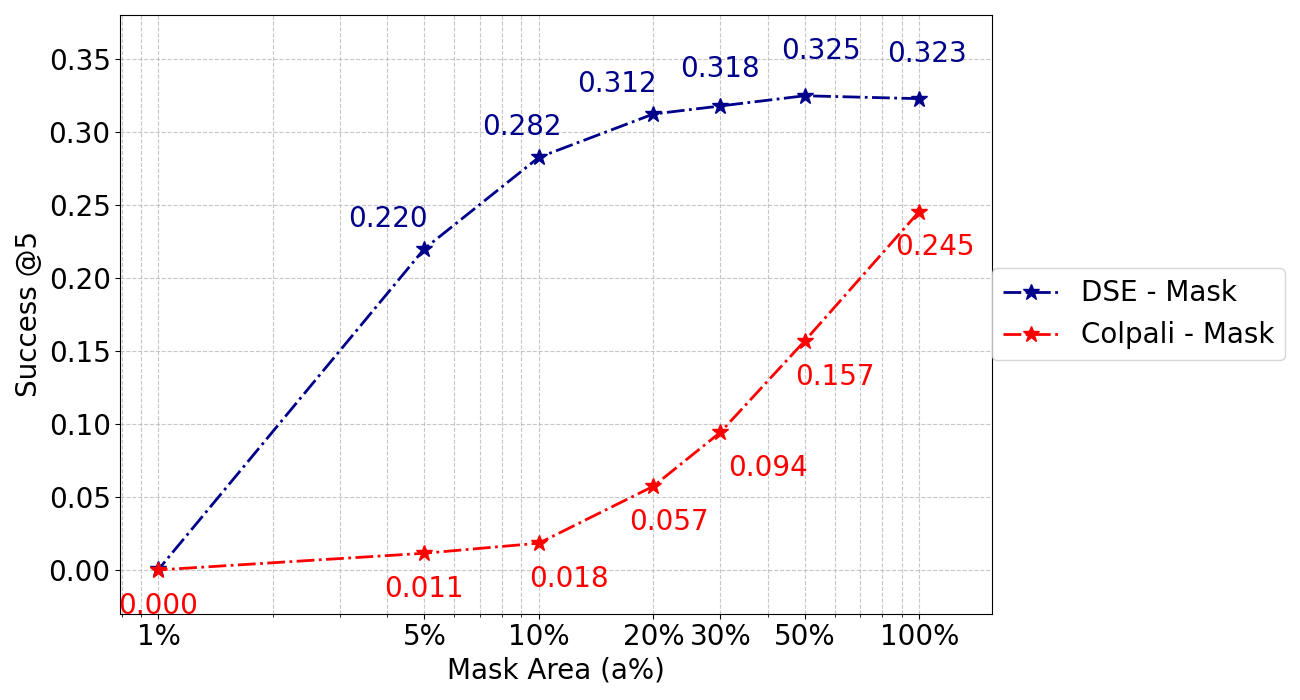}
	
	\caption{Impact of mask area $a$ on attack effectiveness over in-domain queries. Lower percentages of mask area $a$ result in less visual corruption of the document screenshot. \vspace{-10pt}}
	\label{Ablation:mask_percentage}
\end{figure}

\begin{figure*}[t]
	\centering
	%\resizebox{\textwidth}{!}{%
		\begin{tabular}{ccc}
			
			\includegraphics[width=0.285\textwidth]{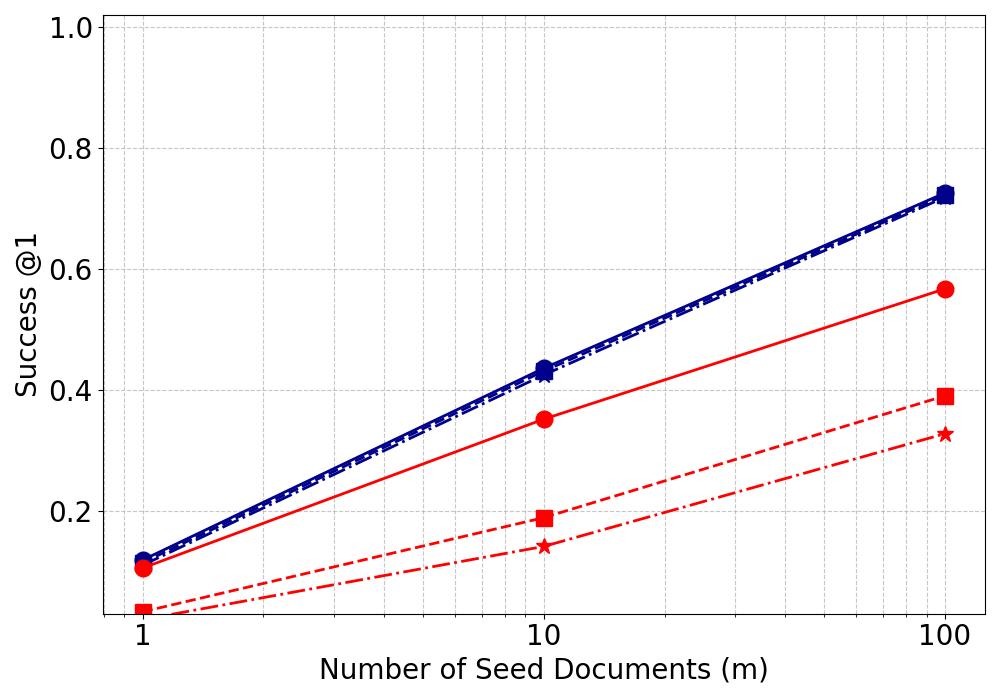}& 
			\includegraphics[width=0.285\textwidth]{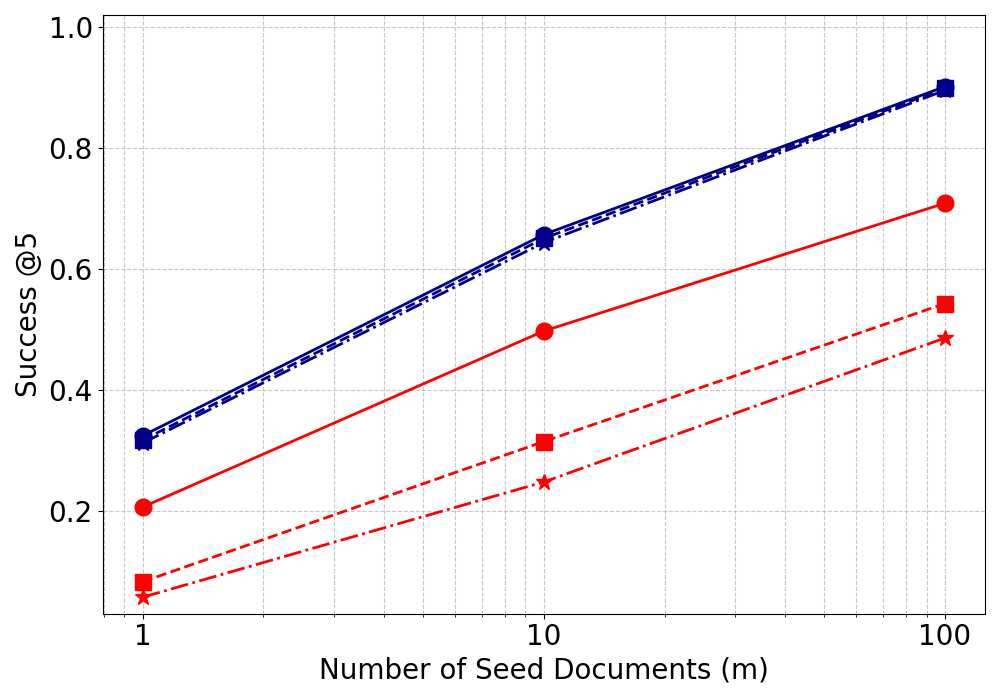} &
			\includegraphics[width=0.36\textwidth]{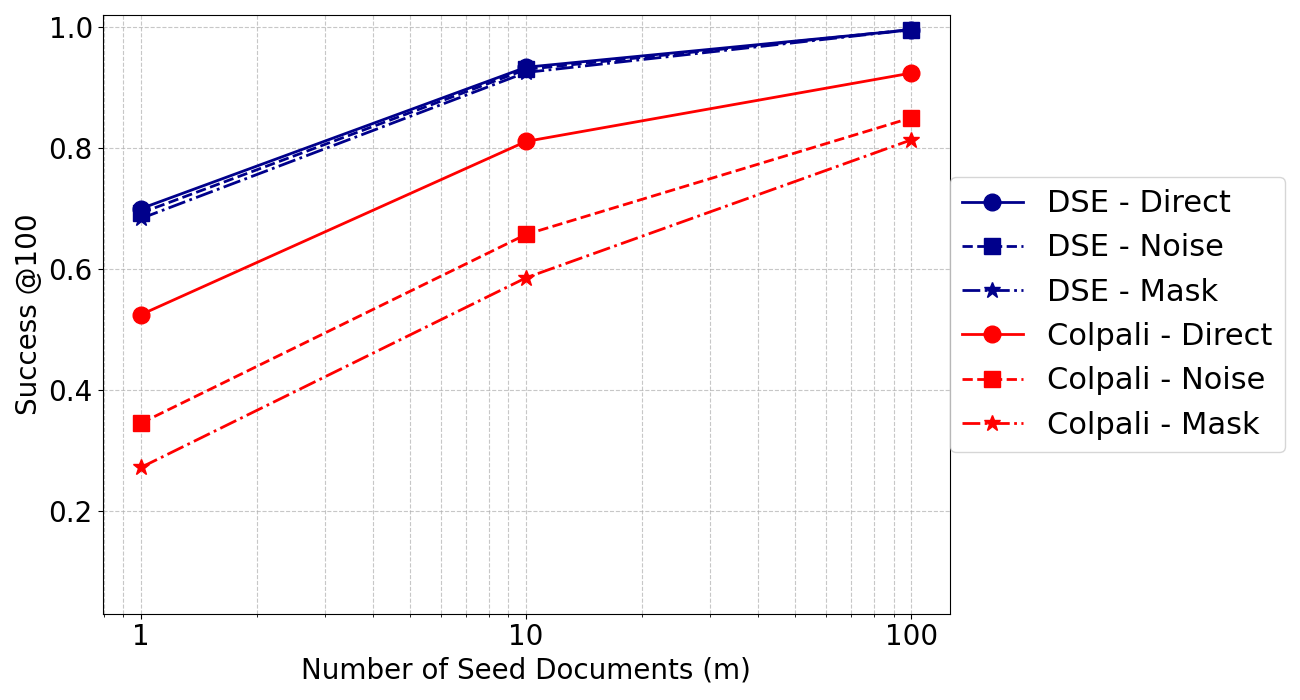} 
			
		\end{tabular}\vspace{-8pt}
		%		%}
	\caption{Impact of the number of injected documents \textbf{m} on on attack effectiveness over in-domain queries. We report results for increasing rank cut-offs: k=1 (left), k=55 (middle), k=100 (right).
	}
	\label{Ablation:k}
\end{figure*}

\begin{table*}[t]
	\centering
		\caption{Results in terms of success@5 obtained when attacking out-of-domain datasets.}\vspace{-5pt}
	\resizebox{.98\textwidth}{!}{%
		\begin{tabular}{l|c|c|cccccccccc}
			\toprule
			%\textbf{Visdore}
			Model
			&$m$
			&Method
			& Arxiv
			& Docv 
			& Infov
			& Tabfquad
			& Tatdqa
			& ShiftPr 
			& SynthAI 
			& SynthEng 
			& SynthGov
			& SyntHth  \\
			\midrule
			\multirow{9}{*}{\textbf{DSE}}  & 	\multirow{3}{*}{\textbf{1}}  & 
			%			& \multicolumn{10}{c}{\textbf{m=1}} \\
			%			\cline{2-11}
			Direct ($p=100\%$)& 0.0260 & 0.6918 & 0.5283 & 0.0250 & 0.0006 & 0.0100 & 0.0400 & 0.0600 & 0.0700 & 0.0300 \\
			&& Noise ($p=100\%$)& 0.0260 & 0.6763 & 0.5020 & 0.0214 & 0.0006 & 0.0100 & 0.0300 & 0.0600 & 0.0600 & 0.0300 \\
			&& Mask ($a=20\%$)& 0.0260 & 0.6696 & 0.4960 & 0.0250 & 0.0006 & 0.0000 & 0.0300 & 0.0600 & 0.0600 & 0.0300 \\
			
			%			& \multicolumn{10}{c}{\textbf{m=10}} \\
			\cline{2-13}
			&	\multirow{3}{*}{\textbf{10}}  &  Direct ($p=100\%$)& 0.0960 & 0.8337 & 0.7287 & 0.0536 & 0.0061 & 0.0100 & 0.2400 & 0.1700 & 0.2200 & 0.2000\\
			&& Noise ($p=100\%$)& 0.0920 & 0.8160 & 0.7146 & 0.0536 & 0.0043 & 0.0100 & 0.2300 & 0.1700 & 0.2000 & 0.1700\\
			&& Mask ($a=20\%$)& 0.0900 & 0.8071 & 0.7065 & 0.0536 & 0.0043 & 0.0100 & 0.2300 & 0.1600 & 0.2000 & 0.1800 \\
			
			%			& \multicolumn{10}{c}{\textbf{m=100}} \\
			\cline{2-13}
			&	\multirow{3}{*}{\textbf{100}} & Direct ($p=100\%$)& 0.1320 & 0.8936 & 0.8522 & 0.2786 & 0.0279 & 0.0700 & 0.5400 & 0.3800 & 0.4200 & 0.4200\\
			&& Noise ($p=100\%$)& 0.1280 & 0.8825 & 0.8462 & 0.2786 & 0.0261 & 0.0700 & 0.5300 & 0.3700 & 0.4200 & 0.4200 \\
			&& Mask ($a=20\%$)& 0.1260 & 0.8847 & 0.8381 & 0.2750 & 0.0255 & 0.0700 & 0.5300 & 0.3600 & 0.4100 & 0.4200 \\
			
			\bottomrule\toprule
			
			\multirow{9}{*}{\textbf{ColPali}}  & 	\multirow{3}{*}{\textbf{1}}  & 
			%			& \multicolumn{10}{c}{\textbf{m=1}} \\
			%			\cline{2-11}
			Direct ($p=100\%$)& 0.0660 & 0.2018 & 0.3664 & 0.1536 & 0.0061 & 0.0000 & 0.1200 & 0.1200 & 0.0500 & 0.0500\\
			&& Noise ($p=100\%$)& 0.0420 & 0.1463 & 0.2773 & 0.1000 & 0.0024 & 0.0000 & 0.0600 & 0.0600 & 0.0300 & 0.0600  \\
			&& Mask ($a=20\%$)& 0.0420 & 0.1153 & 0.2105 & 0.0929 & 0.0012 & 0.0000 & 0.0300 & 0.0400 & 0.0500 & 0.0400 \\
			
			%			& \multicolumn{10}{c}{\textbf{m=10}} \\
			\cline{2-13}
			&	\multirow{3}{*}{\textbf{10}}  &  Direct ($p=100\%$)& 0.1900 & 0.3570 & 0.5567 & 0.2679 & 0.0085 & 0.0100 & 0.2200 & 0.2300 & 0.1600 & 0.1300\\
			&& Noise ($p=100\%$)& 0.1260 & 0.2772 & 0.4555 & 0.2107 & 0.0055 & 0.0000 & 0.1400 & 0.1500 & 0.0800 & 0.1500 \\
			&& Mask ($a=20\%$)& 0.0920 & 0.2262 & 0.3866 & 0.1286 & 0.0043 & 0.0000 & 0.1200 & 0.1000 & 0.0700 & 0.0700 \\
			
			%			& \multicolumn{10}{c}{\textbf{m=100}} \\
			\cline{2-13}
			&	\multirow{3}{*}{\textbf{100}} & Direct ($p=100\%$)& 0.2500 & 0.5233 & 0.6883 & 0.3964 & 0.0194 & 0.0100 & 0.3000 & 0.3800 & 0.2900 & 0.2900 \\
			&& Noise ($p=100\%$)& 0.1640 & 0.4435 & 0.6113 & 0.3393 & 0.0091 & 0.0200 & 0.2300 & 0.2600 & 0.1900 & 0.2300 \\
			&& Mask ($a=20\%$)& 0.1540 & 0.4035 & 0.5688 & 0.2750 & 0.0079 & 0.0000 & 0.1700 & 0.2300 & 0.1600 & 0.1800 \\
			
			\bottomrule
			
		\end{tabular}
	}
	\label{tab:results_visdore_dse}\vspace{-4pt}
\end{table*}

Interestingly, contrary to when attacking a predefined set of target queries, ColPali appears more robust than DSE when attacking a large set of in-domain queries. Furthermore, while different attack methods exhibit similar effectiveness on DSE, direct optimisation demonstrates the best attack effectiveness on ColPali. This suggests that directly modifying pixel values is more effective for attacking ColPali compared to other attack methods.

We explore the trade-off between attack effectiveness and fidelity for this setting in Figures~\ref{Ablation:grad_percentage} and \ref{Ablation:mask_percentage}. Due to the large number of ablation studies and limited computational resources, we optimise a single seed screenshot document instead of 100. However, the low standard deviations reported in Table~\ref{tab:results_nq} suggest this choice is unlikely to significantly affect the generalisability of the observations presented next.

Attacking VLM-based retrievers on a large set of unseen in-domain queries is more challenging than targeting a focused set of predefined queries. Achieving high attack effectiveness in this broader scenario requires increasing the optimised gradient percentage $p$ or the mask area $a$. However, these adjustments reduce attack fidelity by producing more visibly altered documents.

We now explore the possibility of injecting more than one adversarial document screenshot ($m > 1$) into the corpus. For this, we follow previous corpus poisoning studies~\cite{zhong-etal-2023-poisoning, zhuang2024doesvec2textposenew}, which use k-means clustering to group training queries and generate one adversarial screenshot for each cluster. Specifically, we use the DSE model to encode Wiki-SS-NQ training queries into vector representations and perform k-means clustering on these vectors for $m = 1$, $10$,  $100$. Each adversarial screenshot however is generated starting from the same seed document. 
The results are presented in Figure~\ref{Ablation:k}. Injecting more adversarial document screenshots significantly increases attack effectiveness for all methods and models. This result is expected, as having more adversarial screenshots in the corpus increases the likelihood of a query retrieving at least one of them.

\subsection{Results of Attacking out-Domain Datasets} \label{sec_results_outofdomain_queries}
Finally, we evaluate the attacks across the out-of-domain setting. For this, the seed documents were selected from the Wiki-SS-NQ dataset, the attack then used Wiki-SS-NQ training queries to produce the optimised documents, which were then injected into the Vidore benchmark datasets. For this experiment we used one seed document screenshot only for $m=1, 10, 100$, due to the high computational costs involved when considering more seed documents. We report the attack success@5 across all Vidore datasets in Table~\ref{tab:results_visdore_dse}.

%Results are reported in Table~\ref{tab:results_visdore_dse}. 
The adversarial documents trained on Wiki-SS-NQ exhibit good attack generalisation to some out-of-domain datasets. For instance, high attack effectiveness is achieved for the Docv and Infov datasets, even when injecting a single adversarial screenshot ($m=1$). When injecting more adversarial screenshots ($m=100$), most datasets become vulnerable to the attack. However, exceptions include the Tatdqa and ShiftPr datasets, where success@5 remains low despite large values of $m$.
This reduced attack effectiveness may be attributed to the significant domain gap between these datasets and Wiki-SS-NQ. For example, ShiftPr consists of analysts document screenshots in French along with French-language queries, while Wiki-SS-NQ contains English Wikipedia page screenshots and English queries. Despite this, our results demonstrate that the proposed attack methods have the potential to generalise across many out-of-domain datasets, showcasing their broader applicability.

\section{Discussion and Limitations}
\label{sec_discussion}

%Our findings demonstrate the susceptibility of VLM-based document retrievers to attacks that maliciously manipulate documents, enabling manipulated documents to appear more frequently in top-ranking positions. Such methods could be used in corpus poisoning attacks or for black-hat SEO. Through an in-depth analysis of key properties and levers across queries of varying difficulty, as well as several ablation studies, we provided insights into which attacks and configurations are most effective for different attacker objectives. Overall, we found that Colpali is generally less susceptible to our attacks than DSE. 
\subsubsection*{Practical Applicability of Attacks for Corpus Poisoning}
In a typical corpus poisoning attack, the goal is to harm the performance of the retrieval system, degrading user trust. Attack fidelity, i.e. the degree to which a user might be able to identify that a document has been manipulated, is often not of key importance for corpus poisoning. The attacker might want to ensure that the search engine performance is degraded across a specific set of queries, e.g., against queries referring to a specific product brand. In this case, our results show that attacks can be carried out with full success (Table~\ref{tab:ablation_seed_documents_first_experiment}).

On the other hand, the attacker might be interested to detriment the performance of the search engine across any query it receives. If the attacker has knowledge of both the corpus on which retrieval takes place and the overall types of queries the search engine receives, e.g., through a query log, our results suggest that all three attacks are highly effective against DSE, especially for large percentages of gradients or large mask sizes (Section~\ref{sec_results_indomain_queries}) -- settings that are compatible with corpus poisoning as attack fidelity is often not important. However, noise optimisation is generally not an effective attack in this context for ColPali.

Finally, if the attacker still wants  to detriment the performance of the search engine across any query, but has no knowledge of the retrieval corpus and of the type of queries the search engine typically serves, then our results suggest that while these attacks can, in some instances, generalise out-of-domain (e.g., Docv), they remain largely ineffective for other domains (e.g., ShiftPr). % -- see Section~\ref{sec_results_outofdomain_queries}.

\subsubsection*{Practical Applicability of Attacks for SEO}
Next we consider the case of a black-hat SEO attack. 
In this scenario, effectiveness remains crucial, but fidelity also becomes paramount. Unlike an attack designed to degrade system performance and undermine user trust, the objective here is to boost the ranking of a target document and ensure user engagement. Often, such SEO efforts focus on a small set of queries, similar to the settings used in Figures~\ref{Ablation:grad_percentage_exp1} and \ref{Ablation:mask_percentage_exp1}. In this context we found that our attack methods can achieve high effectiveness while maintaining close faithfulness to the original document that has been used in the adversarial optimisation (high attack fidelity). 

When attempting to influence a large set of previously unseen in-domain queries, no attack demonstrates high effectiveness in high-fidelity scenarios (for instance, with a low percentage of gradient optimisation or small mask sizes -- see Figures~\ref{Ablation:grad_percentage} and~\ref{Ablation:mask_percentage}). However, these circumstances are unlikely to be of interest for SEO. 

As previously noted, a key consideration in a black-hat SEO attack is its fidelity. From visual inspection of the documents produced by the attacks (see Figures~\ref{direct_optimization}-\ref{mask_optimization} for examples), we observed that the noise optimisation attack typically produces the least noticeable distortion. Although the documents appear blurry -- particularly at higher gradient percentages -- this blurriness may be mistaken by users for poor webpage buffering or slow loading, rather than deliberate tampering. In contrast, other attacks displaying concentrated noise are far more conspicuous.

\subsubsection*{Limitations}
We acknowledge several limitations in our investigation. In the empirical validation of the attacks in the case of manipulations to a single seed document ($m=1$), we conducted 100 independent experiments, each involving a different seed document. We deliberately sampled documents unlikely to be retrieved by the search engine, making them more challenging to promote to the top of the ranking. However, when examining the manipulation of multiple copies of a seed document ($m=10$ or $m=100$), we tested only one seed document due to prohibitively high computational costs. While in principle this may reduce the generalisability of our findings to other seed documents, our results for $m=1$ were highly consistent across multiple seed documents, particularly for DSE (Table~\ref{tab:results_nq}, note the low standard deviation values).
These stable outcomes provide confidence in the broader applicability of the results obtained with the single representative seed document.

Our findings show that VLM-based document retrievers are susceptible to the three attacks examined; however, defence strategies against data poisoning attacks~\cite{steinhardt2017certified}, e.g., gradient shaping~\cite{hong2020effectiveness}, might prove effective to counteract these attacks. While we did not investigate the impact of such defences here, it is worth noting that they often reduce the efficacy of attacks, albeit at the cost of overall search performance.

Finally, our methods are white-box attacks, requiring access to the retriever's model weights to calculate gradients for updating pixels. This means that if the retrievers are only accessible via API calls, our methods would not be applicable. An interesting direction for future work could be the development of black-box attack methods for document screenshot retrievers.

%Practical limitation of noise injection into the web pages

\section{Conclusion}
In this paper we investigated the susceptibility of VLM-based document retrievers to attacks that maliciously manipulate documents. VLM-based document retrievers have been shown effective across several retrieval tasks and provide the added advantage of simplifying the indexing and retrieval pipeline.

Our findings demonstrate that VLM-based document retrievers are susceptible to attacks, enabling manipulated documents to appear more frequently in top-ranking positions. Such methods could be used in corpus poisoning attacks or for black-hat SEO. Through an in-depth analysis of key properties and levers across queries of varying difficulty, as well as several ablation studies, we provided insights into which attacks and configurations are most effective for different attacker objectives. Overall, we found that ColPali is generally less susceptible to our attacks than DSE, an advantage that we impute to the more complex retrieval mechanism of ColPali. This study contributes to advancing our understanding of the capabilities and robustness of VLM-based document retrievers. 

Code to reproduce the results presented in this paper is available at \url{https://github.com/ielab/dsr-poisoning}.

%\begin{acks}
%	...
%\end{acks}
\balance
\bibliographystyle{ACM-Reference-Format}
\bibliography{bibliography}


\begin{thebibliography}{33}

%%% the Web address.

\ifx \showCODEN    \undefined \def \showCODEN     #1{\unskip}     \fi
\ifx \showDOI      \undefined \def \showDOI       #1{#1}\fi
\ifx \showISBNx    \undefined \def \showISBNx     #1{\unskip}     \fi
\ifx \showISBNxiii \undefined \def \showISBNxiii  #1{\unskip}     \fi
\ifx \showISSN     \undefined \def \showISSN      #1{\unskip}     \fi
\ifx \showLCCN     \undefined \def \showLCCN      #1{\unskip}     \fi
\ifx \shownote     \undefined \def \shownote      #1{#1}          \fi
\ifx \showarticletitle \undefined \def \showarticletitle #1{#1}   \fi
\ifx \showURL      \undefined \def \showURL       {\relax}        \fi
% invisible to TeX
\providecommand\bibfield[2]{#2}
\providecommand\bibinfo[2]{#2}
\providecommand\natexlab[1]{#1}
\providecommand\showeprint[2][]{arXiv:#2}

\bibitem[Castillo et~al\mbox{.}(2011)]%
        {castillo2011adversarial}
\bibfield{author}{\bibinfo{person}{Carlos Castillo}, \bibinfo{person}{Brian~D
  Davison}, {et~al\mbox{.}}} \bibinfo{year}{2011}\natexlab{}.
\newblock \showarticletitle{Adversarial web search}.
\newblock \bibinfo{journal}{\emph{Foundations and trends{\textregistered} in
  information retrieval}} \bibinfo{volume}{4}, \bibinfo{number}{5}
  (\bibinfo{year}{2011}), \bibinfo{pages}{377--486}.
\newblock


\bibitem[Cho et~al\mbox{.}(2024)]%
        {cho2024m3docrag}
\bibfield{author}{\bibinfo{person}{Jaemin Cho}, \bibinfo{person}{Debanjan
  Mahata}, \bibinfo{person}{Ozan Irsoy}, \bibinfo{person}{Yujie He}, {and}
  \bibinfo{person}{Mohit Bansal}.} \bibinfo{year}{2024}\natexlab{}.
\newblock \showarticletitle{M3docrag: Multi-modal retrieval is what you need
  for multi-page multi-document understanding}.
\newblock \bibinfo{journal}{\emph{arXiv preprint arXiv:2411.04952}}
  (\bibinfo{year}{2024}).
\newblock


\bibitem[Dao et~al\mbox{.}(2022)]%
        {flashattention}
\bibfield{author}{\bibinfo{person}{Tri Dao}, \bibinfo{person}{Daniel~Y. Fu},
  \bibinfo{person}{Stefano Ermon}, \bibinfo{person}{Atri Rudra}, {and}
  \bibinfo{person}{Christopher R\'{e}}.} \bibinfo{year}{2022}\natexlab{}.
\newblock \showarticletitle{FLASHATTENTION: fast and memory-efficient exact
  attention with IO-awareness}. In \bibinfo{booktitle}{\emph{Proceedings of the
  36th International Conference on Neural Information Processing Systems}} (New
  Orleans, LA, USA) \emph{(\bibinfo{series}{NIPS '22})}.
  \bibinfo{publisher}{Curran Associates Inc.}, \bibinfo{address}{Red Hook, NY,
  USA}, Article \bibinfo{articleno}{1189}, \bibinfo{numpages}{16}~pages.
\newblock
\showISBNx{9781713871088}


\bibitem[Das et~al\mbox{.}(2024)]%
        {das2024security}
\bibfield{author}{\bibinfo{person}{Badhan~Chandra Das}, \bibinfo{person}{M~Hadi
  Amini}, {and} \bibinfo{person}{Yanzhao Wu}.} \bibinfo{year}{2024}\natexlab{}.
\newblock \showarticletitle{Security and privacy challenges of large language
  models: A survey}.
\newblock \bibinfo{journal}{\emph{Comput. Surveys}} (\bibinfo{year}{2024}).
\newblock


\bibitem[Dong et~al\mbox{.}(2017)]%
        {Dong2017mifgsm}
\bibfield{author}{\bibinfo{person}{Yinpeng Dong}, \bibinfo{person}{Fangzhou
  Liao}, \bibinfo{person}{Tianyu Pang}, \bibinfo{person}{Hang Su},
  \bibinfo{person}{Jun Zhu}, \bibinfo{person}{Xiaolin Hu}, {and}
  \bibinfo{person}{Jianguo Li}.} \bibinfo{year}{2017}\natexlab{}.
\newblock \showarticletitle{Boosting Adversarial Attacks with Momentum}.
\newblock \bibinfo{journal}{\emph{Proceedings of the 31st IEEE/CVF Conference
  on Computer Vision and Pattern Recognition}} (\bibinfo{year}{2017}),
  \bibinfo{pages}{9185--9193}.
\newblock
\urldef\tempurl%
\url{https://api.semanticscholar.org/CorpusID:4119221}
\showURL{%
\tempurl}


\bibitem[Ebrahimi et~al\mbox{.}(2018)]%
        {ebrahimi2018hotflip}
\bibfield{author}{\bibinfo{person}{Javid Ebrahimi}, \bibinfo{person}{Anyi Rao},
  \bibinfo{person}{Daniel Lowd}, {and} \bibinfo{person}{Dejing Dou}.}
  \bibinfo{year}{2018}\natexlab{}.
\newblock \showarticletitle{HotFlip: White-Box Adversarial Examples for Text
  Classification}. In \bibinfo{booktitle}{\emph{Proceedings of the 56th Annual
  Meeting of the Association for Computational Linguistics (Volume 2: Short
  Papers)}}. \bibinfo{pages}{31--36}.
\newblock


\bibitem[Faysse et~al\mbox{.}(2024)]%
        {faysse2024colpali}
\bibfield{author}{\bibinfo{person}{Manuel Faysse}, \bibinfo{person}{Hugues
  Sibille}, \bibinfo{person}{Tony Wu}, \bibinfo{person}{Bilel Omrani},
  \bibinfo{person}{Gautier Viaud}, \bibinfo{person}{C{\'e}line Hudelot}, {and}
  \bibinfo{person}{Pierre Colombo}.} \bibinfo{year}{2024}\natexlab{}.
\newblock \showarticletitle{Colpali: Efficient document retrieval with vision
  language models}.
\newblock \bibinfo{journal}{\emph{arXiv preprint arXiv:2407.01449}}
  (\bibinfo{year}{2024}).
\newblock


\bibitem[Gao et~al\mbox{.}(2023)]%
        {tevatron}
\bibfield{author}{\bibinfo{person}{Luyu Gao}, \bibinfo{person}{Xueguang Ma},
  \bibinfo{person}{Jimmy Lin}, {and} \bibinfo{person}{Jamie Callan}.}
  \bibinfo{year}{2023}\natexlab{}.
\newblock \showarticletitle{Tevatron: An Efficient and Flexible Toolkit for
  Neural Retrieval}. In \bibinfo{booktitle}{\emph{Proceedings of the 46th
  International ACM SIGIR Conference on Research and Development in Information
  Retrieval}} (Taipei, Taiwan) \emph{(\bibinfo{series}{SIGIR '23})}.
  \bibinfo{publisher}{Association for Computing Machinery},
  \bibinfo{address}{New York, NY, USA}, \bibinfo{pages}{3120--3124}.
\newblock
\showISBNx{9781450394086}
\urldef\tempurl%
\url{https://doi.org/10.1145/3539618.3591805}
\showDOI{\tempurl}


\bibitem[Goodfellow et~al\mbox{.}(2015)]%
        {Goodfellow2015fgsm}
\bibfield{author}{\bibinfo{person}{Ian~J. Goodfellow},
  \bibinfo{person}{Jonathon Shlens}, {and} \bibinfo{person}{Christian
  Szegedy}.} \bibinfo{year}{2015}\natexlab{}.
\newblock \showarticletitle{Explaining and Harnessing Adversarial Examples}. In
  \bibinfo{booktitle}{\emph{Proceedings of the 3rd International Conference on
  Learning Representations (ICLR)}}, \bibfield{editor}{\bibinfo{person}{Yoshua
  Bengio} {and} \bibinfo{person}{Yann LeCun}} (Eds.).
\newblock
\urldef\tempurl%
\url{http://arxiv.org/abs/1412.6572}
\showURL{%
\tempurl}


\bibitem[Hong et~al\mbox{.}(2020)]%
        {hong2020effectiveness}
\bibfield{author}{\bibinfo{person}{Sanghyun Hong}, \bibinfo{person}{Varun
  Chandrasekaran}, \bibinfo{person}{Yi{\u{g}}itcan Kaya},
  \bibinfo{person}{Tudor Dumitra{\c{s}}}, {and} \bibinfo{person}{Nicolas
  Papernot}.} \bibinfo{year}{2020}\natexlab{}.
\newblock \showarticletitle{On the effectiveness of mitigating data poisoning
  attacks with gradient shaping}.
\newblock \bibinfo{journal}{\emph{arXiv preprint arXiv:2002.11497}}
  (\bibinfo{year}{2020}).
\newblock


\bibitem[Khattab and Zaharia(2020)]%
        {khattab2020colbert}
\bibfield{author}{\bibinfo{person}{Omar Khattab} {and} \bibinfo{person}{Matei
  Zaharia}.} \bibinfo{year}{2020}\natexlab{}.
\newblock \showarticletitle{Colbert: Efficient and effective passage search via
  contextualized late interaction over bert}. In
  \bibinfo{booktitle}{\emph{Proceedings of the 43rd International ACM SIGIR
  conference on research and development in Information Retrieval}}.
  \bibinfo{pages}{39--48}.
\newblock


\bibitem[Kurakin et~al\mbox{.}(2016)]%
        {Kurakin2016ifgsm}
\bibfield{author}{\bibinfo{person}{Alexey Kurakin}, \bibinfo{person}{Ian~J.
  Goodfellow}, {and} \bibinfo{person}{Samy Bengio}.}
  \bibinfo{year}{2016}\natexlab{}.
\newblock \showarticletitle{Adversarial examples in the physical world}.
\newblock \bibinfo{journal}{\emph{ArXiv}}  \bibinfo{volume}{abs/1607.02533}
  (\bibinfo{year}{2016}).
\newblock
\urldef\tempurl%
\url{https://api.semanticscholar.org/CorpusID:1257772}
\showURL{%
\tempurl}


\bibitem[Kwiatkowski et~al\mbox{.}(2019)]%
        {kwiatkowski-etal-2019-natural}
\bibfield{author}{\bibinfo{person}{Tom Kwiatkowski},
  \bibinfo{person}{Jennimaria Palomaki}, \bibinfo{person}{Olivia Redfield},
  \bibinfo{person}{Michael Collins}, \bibinfo{person}{Ankur Parikh},
  \bibinfo{person}{Chris Alberti}, \bibinfo{person}{Danielle Epstein},
  \bibinfo{person}{Illia Polosukhin}, \bibinfo{person}{Jacob Devlin},
  \bibinfo{person}{Kenton Lee}, \bibinfo{person}{Kristina Toutanova},
  \bibinfo{person}{Llion Jones}, \bibinfo{person}{Matthew Kelcey},
  \bibinfo{person}{Ming-Wei Chang}, \bibinfo{person}{Andrew~M. Dai},
  \bibinfo{person}{Jakob Uszkoreit}, \bibinfo{person}{Quoc Le}, {and}
  \bibinfo{person}{Slav Petrov}.} \bibinfo{year}{2019}\natexlab{}.
\newblock \showarticletitle{Natural Questions: A Benchmark for Question
  Answering Research}.
\newblock \bibinfo{journal}{\emph{Transactions of the Association for
  Computational Linguistics}}  \bibinfo{volume}{7} (\bibinfo{year}{2019}),
  \bibinfo{pages}{452--466}.
\newblock
\urldef\tempurl%
\url{https://doi.org/10.1162/tacl_a_00276}
\showDOI{\tempurl}


\bibitem[Li et~al\mbox{.}(2025)]%
        {li2025reproducing}
\bibfield{author}{\bibinfo{person}{Yongkang Li}, \bibinfo{person}{Panagiotis
  Eustratiadis}, {and} \bibinfo{person}{Evangelos Kanoulas}.}
  \bibinfo{year}{2025}\natexlab{}.
\newblock \showarticletitle{Reproducing HotFlip for Corpus Poisoning Attacks in
  Dense Retrieval}.
\newblock \bibinfo{journal}{\emph{arXiv preprint arXiv:2501.04802}}
  (\bibinfo{year}{2025}).
\newblock


\bibitem[Lin et~al\mbox{.}(2020)]%
        {Lin2020Nesterov}
\bibfield{author}{\bibinfo{person}{Jiadong Lin}, \bibinfo{person}{Chuanbiao
  Song}, \bibinfo{person}{Kun He}, \bibinfo{person}{Liwei Wang}, {and}
  \bibinfo{person}{John~E. Hopcroft}.} \bibinfo{year}{2020}\natexlab{}.
\newblock \showarticletitle{Nesterov Accelerated Gradient and Scale Invariance
  for Adversarial Attacks}. In \bibinfo{booktitle}{\emph{Proceeedings of the
  International Conference on Learning Representations (ICLR)}}.
\newblock
\urldef\tempurl%
\url{https://openreview.net/forum?id=SJlHwkBYDH}
\showURL{%
\tempurl}


\bibitem[Liu et~al\mbox{.}(2024)]%
        {liu2024robust}
\bibfield{author}{\bibinfo{person}{Yu-An Liu}, \bibinfo{person}{Ruqing Zhang},
  \bibinfo{person}{Jiafeng Guo}, \bibinfo{person}{Maarten de Rijke},
  \bibinfo{person}{Yixing Fan}, {and} \bibinfo{person}{Xueqi Cheng}.}
  \bibinfo{year}{2024}\natexlab{}.
\newblock \showarticletitle{Robust neural information retrieval: An adversarial
  and out-of-distribution perspective}.
\newblock \bibinfo{journal}{\emph{arXiv preprint arXiv:2407.06992}}
  (\bibinfo{year}{2024}).
\newblock


\bibitem[Long et~al\mbox{.}(2024)]%
        {long2024backdoor}
\bibfield{author}{\bibinfo{person}{Quanyu Long}, \bibinfo{person}{Yue Deng},
  \bibinfo{person}{LeiLei Gan}, \bibinfo{person}{Wenya Wang}, {and}
  \bibinfo{person}{Sinno~Jialin Pan}.} \bibinfo{year}{2024}\natexlab{}.
\newblock \showarticletitle{Backdoor attacks on dense passage retrievers for
  disseminating misinformation}.
\newblock \bibinfo{journal}{\emph{arXiv preprint arXiv:2402.13532}}
  (\bibinfo{year}{2024}).
\newblock


\bibitem[Luan et~al\mbox{.}(2021)]%
        {luan2021sparse}
\bibfield{author}{\bibinfo{person}{Yi Luan}, \bibinfo{person}{Jacob
  Eisenstein}, \bibinfo{person}{Kristina Toutanova}, {and}
  \bibinfo{person}{Michael Collins}.} \bibinfo{year}{2021}\natexlab{}.
\newblock \showarticletitle{Sparse, dense, and attentional representations for
  text retrieval}.
\newblock \bibinfo{journal}{\emph{Transactions of the Association for
  Computational Linguistics}}  \bibinfo{volume}{9} (\bibinfo{year}{2021}),
  \bibinfo{pages}{329--345}.
\newblock


\bibitem[Ma et~al\mbox{.}(2024a)]%
        {ma2024unifying}
\bibfield{author}{\bibinfo{person}{Xueguang Ma}, \bibinfo{person}{Sheng-Chieh
  Lin}, \bibinfo{person}{Minghan Li}, \bibinfo{person}{Wenhu Chen}, {and}
  \bibinfo{person}{Jimmy Lin}.} \bibinfo{year}{2024}\natexlab{a}.
\newblock \showarticletitle{Unifying Multimodal Retrieval via Document
  Screenshot Embedding}. In \bibinfo{booktitle}{\emph{Proceedings of the 2024
  Conference on Empirical Methods in Natural Language Processing}},
  \bibfield{editor}{\bibinfo{person}{Yaser Al-Onaizan}, \bibinfo{person}{Mohit
  Bansal}, {and} \bibinfo{person}{Yun-Nung Chen}} (Eds.).
  \bibinfo{publisher}{Association for Computational Linguistics},
  \bibinfo{address}{Miami, Florida, USA}, \bibinfo{pages}{6492--6505}.
\newblock
\urldef\tempurl%
\url{https://doi.org/10.18653/v1/2024.emnlp-main.373}
\showDOI{\tempurl}


\bibitem[Ma et~al\mbox{.}(2024b)]%
        {ma2024visa}
\bibfield{author}{\bibinfo{person}{Xueguang Ma}, \bibinfo{person}{Shengyao
  Zhuang}, \bibinfo{person}{Bevan Koopman}, \bibinfo{person}{Guido Zuccon},
  \bibinfo{person}{Wenhu Chen}, {and} \bibinfo{person}{Jimmy Lin}.}
  \bibinfo{year}{2024}\natexlab{b}.
\newblock \showarticletitle{VISA: Retrieval Augmented Generation with Visual
  Source Attribution}.
\newblock \bibinfo{journal}{\emph{arXiv preprint arXiv:2412.14457}}
  (\bibinfo{year}{2024}).
\newblock


\bibitem[Morris et~al\mbox{.}(2023)]%
        {morris-etal-2023-text}
\bibfield{author}{\bibinfo{person}{John Morris}, \bibinfo{person}{Volodymyr
  Kuleshov}, \bibinfo{person}{Vitaly Shmatikov}, {and}
  \bibinfo{person}{Alexander Rush}.} \bibinfo{year}{2023}\natexlab{}.
\newblock \showarticletitle{Text Embeddings Reveal (Almost) As Much As Text}.
  In \bibinfo{booktitle}{\emph{Proceedings of the 2023 Conference on Empirical
  Methods in Natural Language Processing}},
  \bibfield{editor}{\bibinfo{person}{Houda Bouamor}, \bibinfo{person}{Juan
  Pino}, {and} \bibinfo{person}{Kalika Bali}} (Eds.).
  \bibinfo{publisher}{Association for Computational Linguistics},
  \bibinfo{address}{Singapore}, \bibinfo{pages}{12448--12460}.
\newblock
\urldef\tempurl%
\url{https://doi.org/10.18653/v1/2023.emnlp-main.765}
\showDOI{\tempurl}


\bibitem[Radford et~al\mbox{.}(2021)]%
        {radford2021learningtransferablevisualmodels}
\bibfield{author}{\bibinfo{person}{Alec Radford}, \bibinfo{person}{Jong~Wook
  Kim}, \bibinfo{person}{Chris Hallacy}, \bibinfo{person}{Aditya Ramesh},
  \bibinfo{person}{Gabriel Goh}, \bibinfo{person}{Sandhini Agarwal},
  \bibinfo{person}{Girish Sastry}, \bibinfo{person}{Amanda Askell},
  \bibinfo{person}{Pamela Mishkin}, \bibinfo{person}{Jack Clark},
  \bibinfo{person}{Gretchen Krueger}, {and} \bibinfo{person}{Ilya Sutskever}.}
  \bibinfo{year}{2021}\natexlab{}.
\newblock \bibinfo{title}{Learning Transferable Visual Models From Natural
  Language Supervision}.
\newblock
\newblock
\showeprint[arxiv]{2103.00020}~[cs.CV]
\urldef\tempurl%
\url{https://arxiv.org/abs/2103.00020}
\showURL{%
\tempurl}


\bibitem[Riedler and Langer(2024)]%
        {riedler2024beyond}
\bibfield{author}{\bibinfo{person}{Monica Riedler} {and}
  \bibinfo{person}{Stefan Langer}.} \bibinfo{year}{2024}\natexlab{}.
\newblock \showarticletitle{Beyond Text: Optimizing RAG with Multimodal Inputs
  for Industrial Applications}.
\newblock \bibinfo{journal}{\emph{arXiv preprint arXiv:2410.21943}}
  (\bibinfo{year}{2024}).
\newblock


\bibitem[Steinhardt et~al\mbox{.}(2017)]%
        {steinhardt2017certified}
\bibfield{author}{\bibinfo{person}{Jacob Steinhardt}, \bibinfo{person}{Pang
  Wei~W Koh}, {and} \bibinfo{person}{Percy~S Liang}.}
  \bibinfo{year}{2017}\natexlab{}.
\newblock \showarticletitle{Certified defenses for data poisoning attacks}.
\newblock \bibinfo{journal}{\emph{Advances in neural information processing
  systems}}  \bibinfo{volume}{30} (\bibinfo{year}{2017}).
\newblock


\bibitem[Tian et~al\mbox{.}(2022)]%
        {tian2022comprehensive}
\bibfield{author}{\bibinfo{person}{Zhiyi Tian}, \bibinfo{person}{Lei Cui},
  \bibinfo{person}{Jie Liang}, {and} \bibinfo{person}{Shui Yu}.}
  \bibinfo{year}{2022}\natexlab{}.
\newblock \showarticletitle{A comprehensive survey on poisoning attacks and
  countermeasures in machine learning}.
\newblock \bibinfo{journal}{\emph{Comput. Surveys}} \bibinfo{volume}{55},
  \bibinfo{number}{8} (\bibinfo{year}{2022}), \bibinfo{pages}{1--35}.
\newblock


\bibitem[Wang and He(2021)]%
        {Xiaosen2021variance}
\bibfield{author}{\bibinfo{person}{Xiaosen Wang} {and} \bibinfo{person}{Kun
  He}.} \bibinfo{year}{2021}\natexlab{}.
\newblock \showarticletitle{Enhancing the Transferability of Adversarial
  Attacks through Variance Tuning}. In \bibinfo{booktitle}{\emph{Proceedings of
  the IEEE/CVF Conference on Computer Vision and Pattern Recognition (CVPR)}}.
  \bibinfo{publisher}{IEEE Computer Society}, \bibinfo{pages}{1924--1933}.
\newblock
\urldef\tempurl%
\url{https://doi.org/10.1109/CVPR46437.2021.00196}
\showDOI{\tempurl}


\bibitem[Wang et~al\mbox{.}(2022)]%
        {wang2022threats}
\bibfield{author}{\bibinfo{person}{Zhibo Wang}, \bibinfo{person}{Jingjing Ma},
  \bibinfo{person}{Xue Wang}, \bibinfo{person}{Jiahui Hu},
  \bibinfo{person}{Zhan Qin}, {and} \bibinfo{person}{Kui Ren}.}
  \bibinfo{year}{2022}\natexlab{}.
\newblock \showarticletitle{Threats to training: A survey of poisoning attacks
  and defenses on machine learning systems}.
\newblock \bibinfo{journal}{\emph{Comput. Surveys}} \bibinfo{volume}{55},
  \bibinfo{number}{7} (\bibinfo{year}{2022}), \bibinfo{pages}{1--36}.
\newblock


\bibitem[Xia et~al\mbox{.}(2024)]%
        {xia2024mmed}
\bibfield{author}{\bibinfo{person}{Peng Xia}, \bibinfo{person}{Kangyu Zhu},
  \bibinfo{person}{Haoran Li}, \bibinfo{person}{Tianze Wang},
  \bibinfo{person}{Weijia Shi}, \bibinfo{person}{Sheng Wang},
  \bibinfo{person}{Linjun Zhang}, \bibinfo{person}{James Zou}, {and}
  \bibinfo{person}{Huaxiu Yao}.} \bibinfo{year}{2024}\natexlab{}.
\newblock \showarticletitle{MMed-RAG: Versatile Multimodal RAG System for
  Medical Vision Language Models}. In \bibinfo{booktitle}{\emph{NeurIPS Safe
  Generative AI Workshop}}.
\newblock


\bibitem[Yu et~al\mbox{.}(2024)]%
        {yu2024visrag}
\bibfield{author}{\bibinfo{person}{Shi Yu}, \bibinfo{person}{Chaoyue Tang},
  \bibinfo{person}{Bokai Xu}, \bibinfo{person}{Junbo Cui},
  \bibinfo{person}{Junhao Ran}, \bibinfo{person}{Yukun Yan},
  \bibinfo{person}{Zhenghao Liu}, \bibinfo{person}{Shuo Wang},
  \bibinfo{person}{Xu Han}, \bibinfo{person}{Zhiyuan Liu}, {et~al\mbox{.}}}
  \bibinfo{year}{2024}\natexlab{}.
\newblock \showarticletitle{Visrag: Vision-based retrieval-augmented generation
  on multi-modality documents}.
\newblock \bibinfo{journal}{\emph{arXiv preprint arXiv:2410.10594}}
  (\bibinfo{year}{2024}).
\newblock


\bibitem[Zhao et~al\mbox{.}(2024)]%
        {zhao2024dense}
\bibfield{author}{\bibinfo{person}{Wayne~Xin Zhao}, \bibinfo{person}{Jing Liu},
  \bibinfo{person}{Ruiyang Ren}, {and} \bibinfo{person}{Ji-Rong Wen}.}
  \bibinfo{year}{2024}\natexlab{}.
\newblock \showarticletitle{Dense text retrieval based on pretrained language
  models: A survey}.
\newblock \bibinfo{journal}{\emph{ACM Transactions on Information Systems}}
  \bibinfo{volume}{42}, \bibinfo{number}{4} (\bibinfo{year}{2024}),
  \bibinfo{pages}{1--60}.
\newblock


\bibitem[Zhong et~al\mbox{.}(2023)]%
        {zhong-etal-2023-poisoning}
\bibfield{author}{\bibinfo{person}{Zexuan Zhong}, \bibinfo{person}{Ziqing
  Huang}, \bibinfo{person}{Alexander Wettig}, {and} \bibinfo{person}{Danqi
  Chen}.} \bibinfo{year}{2023}\natexlab{}.
\newblock \showarticletitle{Poisoning Retrieval Corpora by Injecting
  Adversarial Passages}. In \bibinfo{booktitle}{\emph{Proceedings of the 2023
  Conference on Empirical Methods in Natural Language Processing}},
  \bibfield{editor}{\bibinfo{person}{Houda Bouamor}, \bibinfo{person}{Juan
  Pino}, {and} \bibinfo{person}{Kalika Bali}} (Eds.).
  \bibinfo{publisher}{Association for Computational Linguistics},
  \bibinfo{address}{Singapore}, \bibinfo{pages}{13764--13775}.
\newblock
\urldef\tempurl%
\url{https://doi.org/10.18653/v1/2023.emnlp-main.849}
\showDOI{\tempurl}


\bibitem[Zhuang et~al\mbox{.}(2024b)]%
        {zhuangv2t}
\bibfield{author}{\bibinfo{person}{Shengyao Zhuang}, \bibinfo{person}{Bevan
  Koopman}, \bibinfo{person}{Xiaoran Chu}, {and} \bibinfo{person}{Guido
  Zuccon}.} \bibinfo{year}{2024}\natexlab{b}.
\newblock \showarticletitle{Understanding and Mitigating the Threat of Vec2Text
  to Dense Retrieval Systems}. In \bibinfo{booktitle}{\emph{Proceedings of the
  2024 Annual International ACM SIGIR Conference on Research and Development in
  Information Retrieval in the Asia Pacific Region}} (Tokyo, Japan)
  \emph{(\bibinfo{series}{SIGIR-AP 2024})}. \bibinfo{publisher}{Association for
  Computing Machinery}, \bibinfo{address}{New York, NY, USA},
  \bibinfo{pages}{259--268}.
\newblock
\showISBNx{9798400707247}
\urldef\tempurl%
\url{https://doi.org/10.1145/3673791.3698414}
\showDOI{\tempurl}


\bibitem[Zhuang et~al\mbox{.}(2024a)]%
        {zhuang2024doesvec2textposenew}
\bibfield{author}{\bibinfo{person}{Shengyao Zhuang}, \bibinfo{person}{Bevan
  Koopman}, {and} \bibinfo{person}{Guido Zuccon}.}
  \bibinfo{year}{2024}\natexlab{a}.
\newblock \bibinfo{title}{Does Vec2Text Pose a New Corpus Poisoning Threat?}
\newblock
\newblock
\showeprint[arxiv]{2410.06628}~[cs.IR]
\urldef\tempurl%
\url{https://arxiv.org/abs/2410.06628}
\showURL{%
\tempurl}


\end{thebibliography}

\end{document}